\documentclass[useAMS]{mn2e}
\usepackage{epsfig}
\usepackage{amsmath,amssymb}
\newcommand{\be}{\begin{equation}}    
\newcommand{\ee}{\end{equation}}

\newcommand{\Enstatite}{{\rm MgSiO$_3$}}
\newcommand{\Forsterite}{{\rm Mg$_2$SiO$_4$}}
\newcommand{\Magnetite}{{\rm Fe$_3$O$_4$}}
\newcommand{\Silica}{{\rm SiO$_2$}}
\newcommand{\Alumina}{{\rm Al$_2$O$_3$}}
\newcommand{\Msun}{$\rm M_{\odot}$}

\title[The metal and dust yields of the first massive stars]{The metal and dust yields of the first massive stars}
\author[S. Marassi et al.]{Stefania Marassi$^{1}$\thanks{E-mail:
stefania.marassi@oa-roma.inaf.it}, Raffaella Schneider$^{1}$, Marco Limongi$^{1,3}$, Alessandro Chieffi$^{2}$,
\newauthor
Marco Bocchio$^{1,4}$, Simone Bianchi$^{4}$ \\
$^{1}$INAF/Osservatorio Astronomico di Roma, Via di Frascati 33, 00040 Monteporzio, Italy \\
$^{2}$INAF/IASF, Via Fosso del Cavaliere 100, 00133 Roma, Italy\\
$^{3}$Kavli Institute for the Physics and Mathematics of the Universe, Todai Institutes for Advanced Study,\\
The University of Tokyo, Kashiwa 277-8583, Japan\\
$^{4}$INAF/Osservatorio Astrofisico di Arcetri, Largo Enrico Fermi 5, 50125 Firenze, Italy}
\begin{document}

\date{}

\pagerange{\pageref{firstpage}--\pageref{lastpage}} \pubyear{2015}

\maketitle

\label{firstpage}

\begin{abstract}
We quantify the role of Population (Pop) III core-collapse supernovae (SNe) as the first cosmic
dust polluters. Starting from a homogeneous set of stellar progenitors with masses in the range $\rm [13 - 80]\, M_\odot$,
we find that the mass and composition of newly formed dust depend on the mixing efficiency of the ejecta and the degree
of fallback experienced during the explosion.

For {\it standard} Pop~III SNe, whose explosions are calibrated to reproduce the average elemental abundances of Galactic halo
stars with $\rm [Fe/H] < -2.5$, between 0.18 and 3.1\,$\rm M_\odot$
($0.39 - 1.76\,\rm M_\odot$) of dust can form in uniformly mixed (unmixed) ejecta, and the dominant grain species are silicates.
We also investigate dust formation in the ejecta of {\it faint} Pop~III SN, where the ejecta experience a strong fallback.
By examining a set of models, tailored to minimize the scatter with the abundances of carbon-enhanced Galactic halo stars
with $\rm [Fe/H ] < -4$, we find that amorphous carbon is the only grain
species that forms, with masses in the range $\rm 2.7 \times 10^{-3}-\rm 0.27 \,\rm M_\odot$ ($\rm 7.5 \times 10^{-4} -\rm 0.11 \, M_\odot$)
for uniformly mixed (unmixed) ejecta models.

Finally, for all the models we estimate the amount and composition of dust that survives the passage of the reverse shock,
and find that, depending on circumstellar medium densities, between 3 and 50\% (10 - 80\%) of dust produced by {\it standard}
({\it faint}) Pop~III SNe can contribute to early dust enrichment.
\end{abstract}

\begin{keywords}
stars:low-mass, supernovae: general, ISM: cloud, dust, Galaxy: halo, galaxies:evolution 
\end{keywords}

\section{Introduction}
Dust grains play a fundamental role in the 
evolution of stellar populations at high redshift. 
Population~III stars (hereafter Pop~III), so far undetected, are 
responsible of the first chemical enrichment of the high-redshift 
interstellar medium (ISM). Given the current theoretical limits on the 
mass of the first stars (Hosokawa et al.~2011; Hosokawa et al.~2012), 
early metal enrichment is likely to be mostly driven by the first core-collapse 
supernovae (SNe), with a possible contribution from more massive pair-instability 
supernovae (although this may strongly depend on the poorly constrained tail 
of the stellar initial mass function, Hirano et al.~2014, 2015; Susa et al.~2014). 

Yet, the amount and properties of grains that can be injected in the high-redshift 
ISM and contribute to the enrichment depend on the dust condensation 
efficiencies in SN ejecta and on the destruction suffered by thermal 
and non-thermal sputtering during the passage of the reverse shock 
of the SN, on timescales of $\rm 10^4 yr$ (Bianchi \& Schneider~2007; 
Nozawa et al.~2007). Depending on the density of the circumstellar medium 
where the explosion takes place, the mass fraction of newly formed dust 
that is able to survive ranges between $2\%$ and $20\%$ for circumstellar 
medium densities in the range $10^{-25}<\rho/(\rm gr/cm^3)<10^{-23}$, which 
corresponds to number densities in the range $\rm 0.06 < n_{ISM}/(\rm cm^{-3}) < 6$ 
(Bianchi \& Schneider~2007; Nozawa et al.~2007; Silvia et al.~2010; 
Silvia et al.~2012; Marassi et al.~2014). In addition, the passage 
of the reverse shock significantly alters the grain size distribution 
and modifies the grain cross section, changing the dust cooling 
efficiency (Schneider \& Omukai~2010).

Dust formation in SN ejecta has been investigated following two 
different methods: classical nucleation theory (CNT) and a chemical 
kinetic approach (Cherchneff \& Dwek~2009; Cherchneff \& Dwek~2010; Sarangi \& 
Cherchneff~2013). Theoretical models, developed in the framework 
of CNT, have shown that dust formation can take place in SN
ejecta a few hundred days after the explosions and provide predictions 
on the mass, composition and size distribution of the newly formed grains 
(Kozasa et al.~1989,~1991; Todini \& Ferrara~2001; Nozawa et al.~2003,
~2008,~2010,~2011; Schneider et al.~2004; Bianchi \& Schneider~2007). 
These models predict dust masses $\approx [0.2 - 0.6]\,$M$_{\odot}$ for SN 
progenitors with masses $\rm [12 - 40]\, M_{\odot}$ and metallicities $[0.1 - 1]\, Z_{\odot}$ 
(Schneider et al.~2014), in agreement with the dust mass inferred 
from recent Herschel and ALMA observations of SN 1987A (Matsuura et al.~2011; 
Indebetouw et al.~2014) and Cas A (Barlow et al.~2010). Although in the past 
the applicability of CNT in astrophysical context has been questioned (Donn \& Nuth~1985; 
Cherchneff \& Dwek~2009), mainly due to the assumption of chemical equilibrium 
at nucleation, recently Paquette \& Nuth~(2011) showed that this assumption has a 
lower impact on the grain mass and size distribution than previously 
thought. Similarly, Nozawa \& Kozasa~(2013) demonstrated that CNT 
 is a good approximation in SN ejecta, at least until the 
collisional timescales of the key molecule is much smaller than the 
timescale with which the supersaturation ratio increases.

The goal of the present study is to investigate the role of Pop~III core-collapse SN 
as dust polluters, adopting a homogeneous set of metal-free progenitors with masses in 
the range $[13 - 80] \, \rm M_{\odot}$. Under the assumption that the observed metal poor 
stars likely formed from gas clouds enriched by Pop~III SNe, we use properly calibrated 
SN explosion models (Chieffi \& Limongi 2002) to 
calculate the formation of dust in the ejecta, applying CNT and accounting for the 
process of grain growth. We quantify the impact of ejecta mixing on the final 
dust masses and grain size distributions, analyzing both uniformly mixed models within 
the helium core and unmixed/stratified cases. To obtain a more realistic estimate of the 
dust mass that is able to enrich the ISM, we also consider the effects 
of the reverse shock (Bianchi \& Schneider~2007). 

Motivated by the observed surface elemental abundances of Carbon Enhanced Metal Poor (CEMP)
stars, i.e. stars with an observed overabundance of light elements compared to Fe and - in particular -
with [C/Fe]$>1$ (Beers \& Christlieb 2005), we also investigate dust formation in the ejecta 
of {\it faint} Pop~III SN, where the ejecta experience mixing and fallback 
(Umeda \& Nomoto~2002; Umeda \& Nomoto~2003). Recently, Marassi et al.~(2014) showed that 
dust can be produced in faint Pop~III SN ejecta (see Kochanek~2014 for dust formation 
in solar metallicity faint SNe.)

Here, we extend this previous analysis and we simulate a large set of faint Pop~III SN models, 
searching for a combination of mixing and fallback that provides the best-fit to the observed abundance 
pattern of all currently known C-enhanced hyper-iron-poor stars (Beers \& Christlieb~2005; 
Yong et al.~2013; Keller et al.~2014; Ishigaki et al.~2014). For these faint SN 
models we also explore the impact, on the final dust masses, of ejecta mixing and 
reverse shock.

The plan of the paper is as follows. In Section 2 we briefly summarize the 
Bianchi \& Schneider (2007) dust nucleation model and we describe the 
upgrated molecular network. In Section 3 we illustrate the main features 
of Pop~III SN progenitors that are modeled using FRANEC stellar evolutionary 
code (Limongi \& Chieffi~2006; Limongi \& Chieffi~2012) and their calibration. 
In Section 4 we present the resulting dust yields for Pop~III 
core-collapse SNe. In Section 5 we describe how we construct, using 
the mixing and fallback procedure (Umeda \& Nomoto~2002), faint Pop~III 
ejecta calibrated with the same procedure described in Marassi et al.~(2014). 
In Section 6 we show the results on dust yields obtained in Pop~III faint SN ejecta. 
In Section 7 we discuss all the results and their dependence on the fallback, ejecta 
mixing, and reverse shock and we draw our conclusions. In the Appendix we present our 
final dust data grids that will be available to the community.\footnote{The resulting dust 
and metal yields will be available in electronic format for interested researchers upon request.}

\section{Dust formation model}
To model dust formation in SN ejecta we follow the Bianchi 
\& Schneider~(2007) model, where classical nucleation theory 
in steady state conditions was applied (see Nozawa et al.~2003; 
Bianchi \& Schneider~2007 and references therein). For our calculation, 
we use a previously developed code that has been applied to core-collapse 
(Todini \& Ferrara~2001; Bianchi \& Schneider~2007) and pair-instability SNe 
(Schneider et al.~2004). This theoretical model has allowed to successfully 
reproduce the dust masses observed in SNe and young SN remnants (Schneider et 
al.~2014; Valiante \& Schneider~2014). In Bianchi \& Schneider (2007) the 
chemistry of molecular formation is implemented following Todini \& Ferrara~(2001), 
but relaxing the assumption of steady state. In the gas-phase, the formation 
of carbon oxide (CO) and silicon oxide (SiO) was assumed to be driven by 
radiative association reactions and destroyed by Compton electrons coming from 
radioactive decay of $^{56}$Co. Seven different grain species are formed in SN ejecta: 
amorphous carbon (AC), iron, corundum (\Alumina), magnetite (\Magnetite), enstatite 
(\Enstatite), forsterite (\Forsterite) and quarz (\Silica). In this work, we follow the 
formation of the above grain species assuming that seed clusters are formed 
by a minimum of two monomers, which subsequently grow by accretion of other monomers. 
The accretion process is regulated by the collisional rate of the key species and 
depends on the sticking coefficient (defined as the probability that an atom colliding 
with a grain will stick to it) which we have assumed equal to 1 for all grain species.
A discussion of the dependence of the dust yields on these two parameters can be found
in Bianchi \& Schneider~(2007), where is also possible to find a description of the adopted 
properties of SN dust species to which we refer for more details. 

\begin{table*}
\begin{tabular}{@{}llll}
\hline
TYPE  & REACTION& RATE COEFFICIENT& REFERENCE\\
\hline
\hline
RA1 & ${\rm C+O\rightarrow CO+h\nu}$& k$_{\rm RA1} = 1.58 \times 10^{-17}(\rm T/300)^{0.3}\exp(-1297.4/T)$& Dalgarno et al.~(1990)\\
RA2 & ${\rm Si+O\rightarrow SiO+h\nu}$&k$_{\rm RA2} = 5.52 \times 10^{-18}(\rm T/300)^{0.3}$& UMIST\\
RA3 & ${\rm C+C\rightarrow C_2+h\nu}$&k$_{\rm RA3} = 4.36 \times 10^{-18}(\rm T/300)^{0.3}\exp(-161.3/T)$&Andreazza \& Sigh~(1987)\\
RA4 & ${\rm O+O\rightarrow O_2+h\nu}$ & see text& Babb \& Dalgarno~(1994)\\
\hline
NN1 & ${\rm C+O_2\rightarrow CO+O}$ &k$_{\rm NN1} = 5.56 \times 10^{-11}(\rm T/300)^{0.4}\exp(26.90/T)$&UMIST\\
NN2 & ${\rm O+C_2\rightarrow CO+C}$ &k$_{\rm NN2} = 2.0\times 10^{-10}(\rm T/300)^{-0.1}$&UMIST\\
NN3 & ${\rm C+CO\rightarrow C_2+O}$ &k$_{\rm NN3} = 2.94\times 10^{-11}(\rm T/300)^{0.5}\exp(-58025/T)$&UMIST\\
NN4 & ${\rm O+CO\rightarrow O_2+C}$ & see text& - \\
NN5 & ${\rm Si+CO\rightarrow SiO+C}$&k$_{\rm NN6} = 1.30\times 10^{-9}\exp(\rm -34513/T)$&UMIST\\        
NN6 & ${\rm C+SiO\rightarrow CO+Si}$& k$_{\rm NN8} = 1.00\times 10^{-16}$&UMIST\\
\hline
\end{tabular}
\caption{Molecular processes considered in this work: the rate coefficients are taken from the UMIST database for astrochemistry 2012; 
where the origin of the reaction rates is different the corresponding reference is indicated.}
\label{Tableform}
\end{table*}
\begin{table*}
\begin{tabular}{@{}lll}
\hline
TYPE  & REACTION & W$_{i}$ [eV]\\
\hline
\hline
D1 &${\rm CO + e^{-}\rightarrow C+O +e^{-}}$ &W$_{\rm i, CO}$=125 \\
D2 &${\rm SiO + e^{-}\rightarrow Si+O+e^{-}}$ &W$_{\rm i, SiO}$=110 \\
D3 &${\rm O_2 + e^{-}\rightarrow O+O+e^{-}}$ &W$_{\rm i, O_2}$=125 \\
D4 &${\rm C_2 + e^{-}\rightarrow C+C+e^{-}}$ &W$_{\rm i, C_2}$=125 \\
\hline
\end{tabular}
\caption{Compton Electron destruction reactions and the corresponding mean energy per ion pair $\rm W_{\rm i}$.}    
\label{Tabledestr}
\end{table*}

\subsection{Upgrated Molecular Network}
It is well known that CO and SiO  molecules play a fundamental role in the 
dust formation process: CO formation subtracts C-atoms and limits 
the formation of AC grains. SiO is required to form 
silicates, such as \Forsterite~and \Enstatite. 
Due to the important role of molecules in the dust formation pathway, 
we have enlarged our molecular network taking into account other 
formation/destruction processes involving CO/SiO molecules and 
their interactions with O$_2$ and C$_2$. These molecular processes 
are crucial in subtracting gas-phase elements from the ejecta 
(in particular oxygen and carbon, which are very abundant, see left panel of 
Fig.~\ref{MpMg_MpMet}), that otherwise are free to form dust. 
We follow the evolution  of CO, SiO, O$_2$ and C$_2$ 
which form through different channels (reaction rates), respectively: 
(i) radiative association reactions where the formation of molecules 
takes place through the emission of a photon which carries off the excess 
energy released during the formation process; (ii) bimolecular, neutral-neutral 
reactions that involve molecules and atoms. For these two-body reactions, the formation 
rates of molecules \rm{k(T)} are given by the usual Arrhenius-type 
expression,
\begin{equation}
{\rm k(T)}=\alpha \left[\frac{T}{300 \rm K}\right]^{\beta}\exp\left[\frac{-\gamma}{T}\right] {\rm cm^3}{\rm s^{-1}}
\label{Arrform}
\end{equation}
where T is the temperature of the ejecta gas in K and $\gamma$ the activation energy in K. 
In Table \ref{Tableform} we report the rates expressed in the Arrhenius form according to 
the UMIST database for astrochemistry 2012\footnote{http://udfa.ajmarkwick.net}. This database is a 
compilation of molecular rates that have different origin, some are theoretically 
calculated, others are directly measured in the laboratory. 
Clearly, there is a degree of uncertainty related to the rate calculations: in some 
cases we decided to refer to other rate estimates present in the literature that are 
more robust. This is the case for the radiative association rate coefficient 
of O$_2$ (RA4), for which we have chosen to fit the expression of the rate coefficient, 
as a function of temperature, obtained from theoretical calculations in Babb \& Dalgarno (1994). 
In addition, we found that the neutral backward reaction NN4 is negligible (K. Omukai, private 
communication). 

In standard nucleation theory, dust condensation is described in terms of a 
nucleation current that depends on the abundance of the key species. Hence,
contrary to other studies which adopt a chemical kinetic approach (Cherchneff \& Dwek 2009, 2010;
Sarangi \& Cherchneff 2015), we do not follow the formation of carbon chains as a pathway
to solid carbon clusters. Thermal fragmentation of the chains through collisions with gas particles, in addition
to oxydation reactions similar to NN2, may limit the formation of carbon chains. However, if all C$_2$
were to contribute to carbon dust formation, the estimated carbon dust mass would need to be corrected
upward by $\rm M_{C_2}$.

We take into account the destruction due to Compton electrons coming from 
the $^{56}$Ni decay in the ejecta. As observed in SN 1987A, destruction by Compton electrons 
has a deep impact on the ejecta chemistry. As shown by Woosley et al. (1989), the explosion of SN 1987A 
has produced $^{56}$Ni which decays in $\sim 6$ days into $^{56}$Co; the subsequent 
decay of $^{56}$Co into $^{56}$Fe deposits energy as $\gamma$ rays in the SN ejecta,
powering the observed light curve. In SN 1987A, the emitted light curve is very well 
reproduced if $\rm 0.075 \,M_{\odot}$ mass of $^{56}$Co was ejected during the SN explosion. 

Here we assume molecules to be destroyed by the impact with energetic electrons produced 
by the radioactive decay of $^{56}$Co, with a rate coefficient k$_{\rm d}$ that depends also 
on the mean energy per ion pair W$_i$. According to Woosley et al. (1989), this rate 
coefficient can be estimated as follows: the thermalized $\gamma$-ray energy input 
rate for a given $^{56}$Co mass is given by,
\begin{equation}
\rm L_{\gamma} = 9.54 \times 10^{41} \left(\frac{M_{\rm ^{56}Co}}{0.075 M_{\odot}}\right)
\rm f_{\gamma}(\rm k_{56}) e^{-t/\tau_{56}} {\rm erg/s},
\label{Lgamma}
\end{equation}
\noindent
where $\rm \, 0.075 M_{\odot}$ was the adopted mass of $^{56}$Co produced in the original
Woosley et al. (1989) model, $\rm \tau_{56} = 111.26$ days is the e-folding time of 
$^{56}$Co decay, and the function $\rm f_{\gamma}$ is the deposition function and it is
given by one minus the fraction of energy that escapes in photons in the X-ray and
$\gamma$-ray bands,
\begin{equation}
f_{\gamma}(\rm k_{56}) = 1 - e^{[-\rm k_{56} \phi_0(t_0/t)^2]},
\label{eq:fdep}
\end{equation}
\noindent
where $\rm \phi_0 = \phi(t_0)$ represents the column depth of the SN
at some fiducial time $\rm t_0$ (for $\rm t_0 = 10^6 $s $\rm \phi_0 = 7 \times 10^4$gr/cm$^2$)
and $\rm k_{56} = 0.033$ cm$^2$ gr$^{-1}$ is an average opacity to  $\gamma$ rays
from $^{56}$Co decay. The above quoted values of $\rm k_{56}$ and $\rm \phi_0$ have 
been derived by Woosley et al.~(1989) for  SN 1987A and we will assume these to hold for all 
the explored SN progenitors.
In particular, we assume that once a $\gamma$-ray Compton scatters with electrons, 
it is completely absorbed. Thus, the $\rm L_{\gamma}$ can be considered as the electron 
energy input and the energy transferred to a single gas particle per unit time can 
be computed as,
\begin{equation}
\rm L_{\rm e} = \rm  L_{\gamma}/N_{\rm part}
\end{equation}
\noindent
where $\rm N_{\rm part}$ is the number of gas particles in the ejecta. To compute the 
destruction rate, $\rm k_{d}$, it is necessary to divide $\rm L_{\rm e}$ by the mean energy 
per dissociation, $\rm W_{\rm i}$. For example, to compute the destruction rate of neutral 
CO we divide L$_{\rm e}$ by the mean energy per dissociation W$_{\rm i, CO}$, obtaining 
\begin{equation}
\rm k_{\rm d}(\rm CO) =\rm L_{\rm e}/W_{\rm i}\qquad {\rm s^{-1}}.
\end{equation}
Finally, we assume that all the radioactive energy is deposited uniformly 
in the ejecta. As it will be clear in what follows, the efficiency of molecule formation/destruction
processes depends on the chemical composition and on the thermodynamics of the ejecta.
In Section 4, we give a detailed description of the relevant processes for some selected
SN models.
\begin{figure*}
\hspace{-1.0cm}
\includegraphics[width=7.65cm]{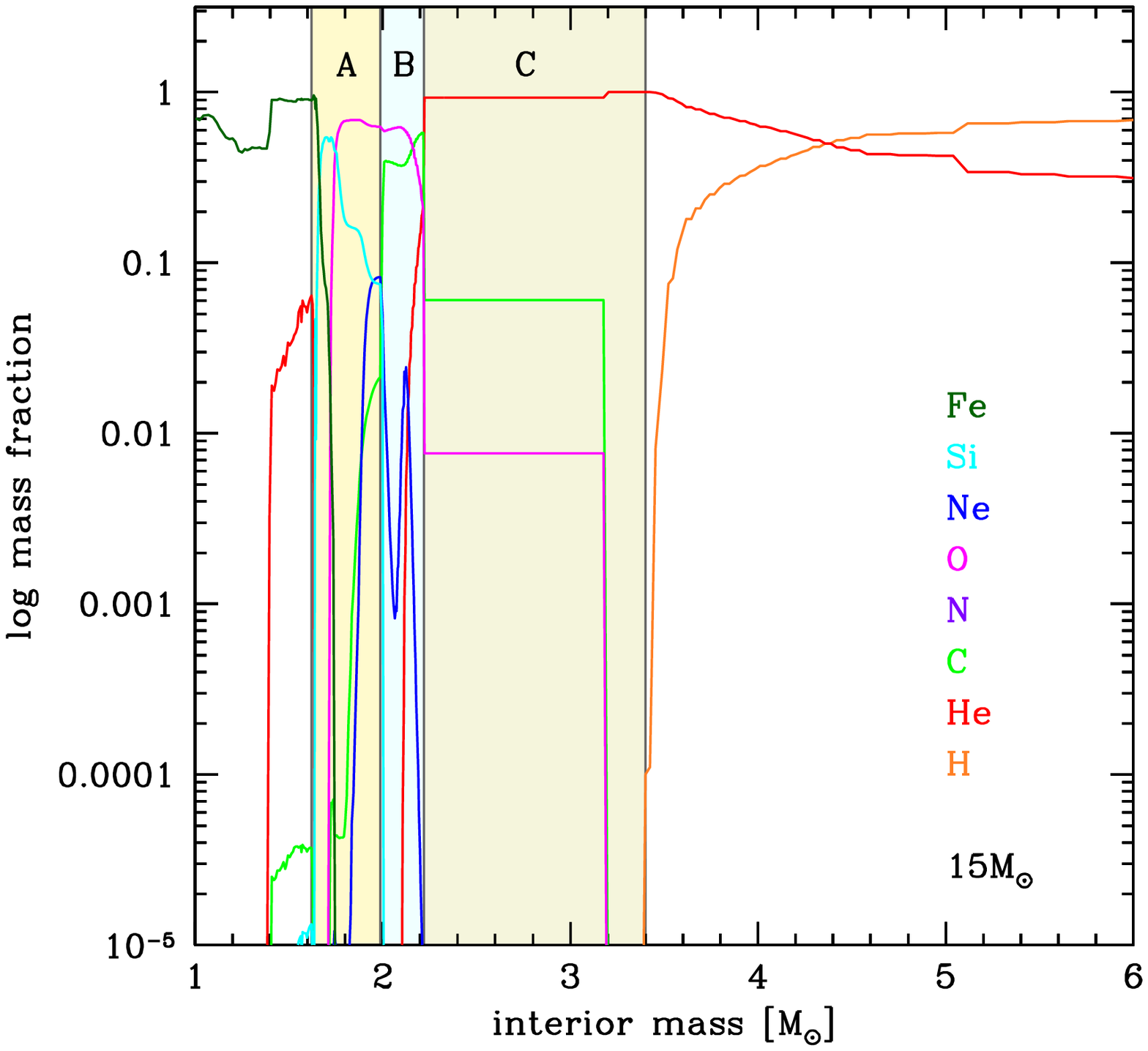}
\hspace{-2.5cm}
\includegraphics[width=7.65cm]{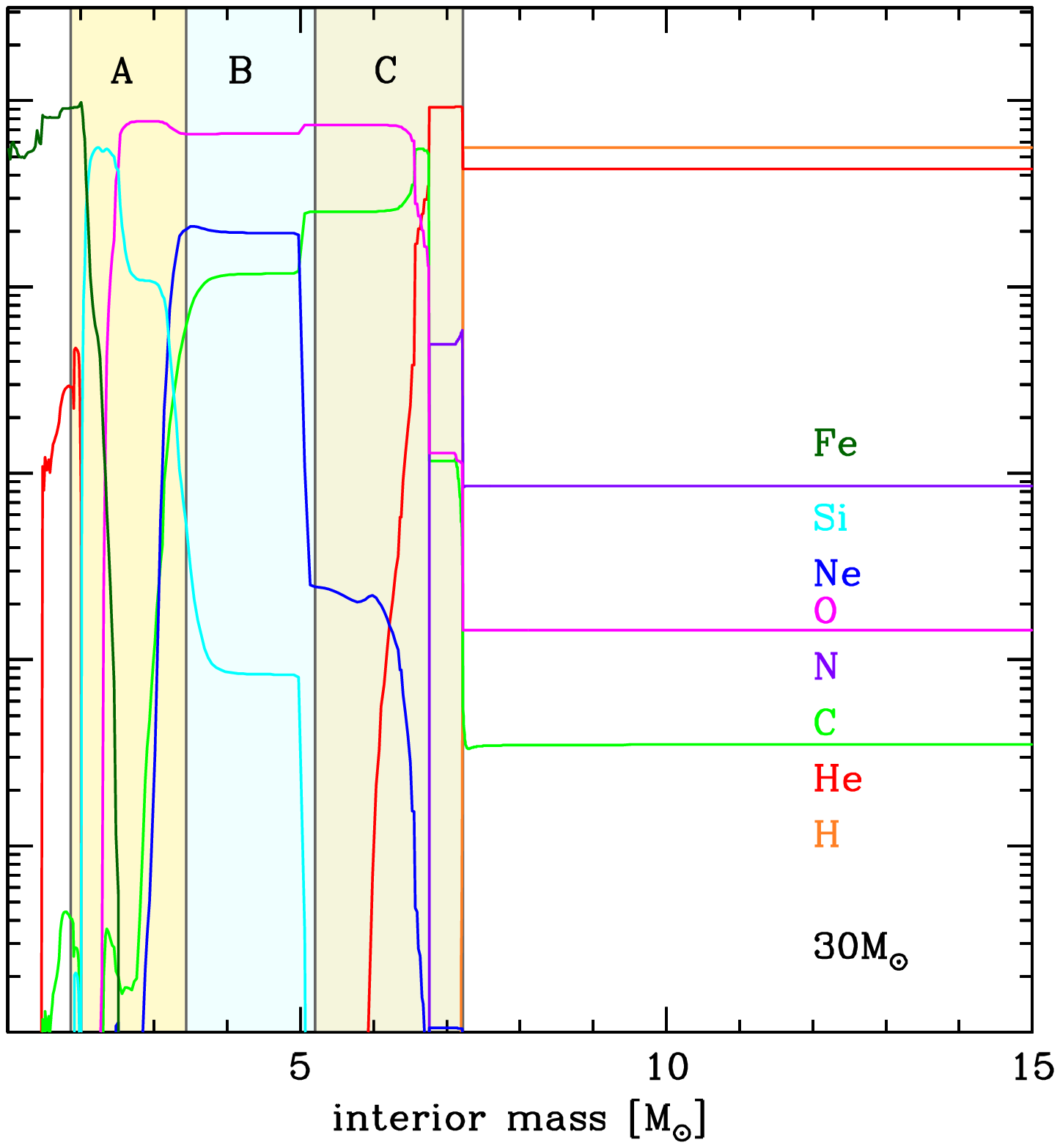}
\hspace{-2.5cm}
\includegraphics[width=7.65cm]{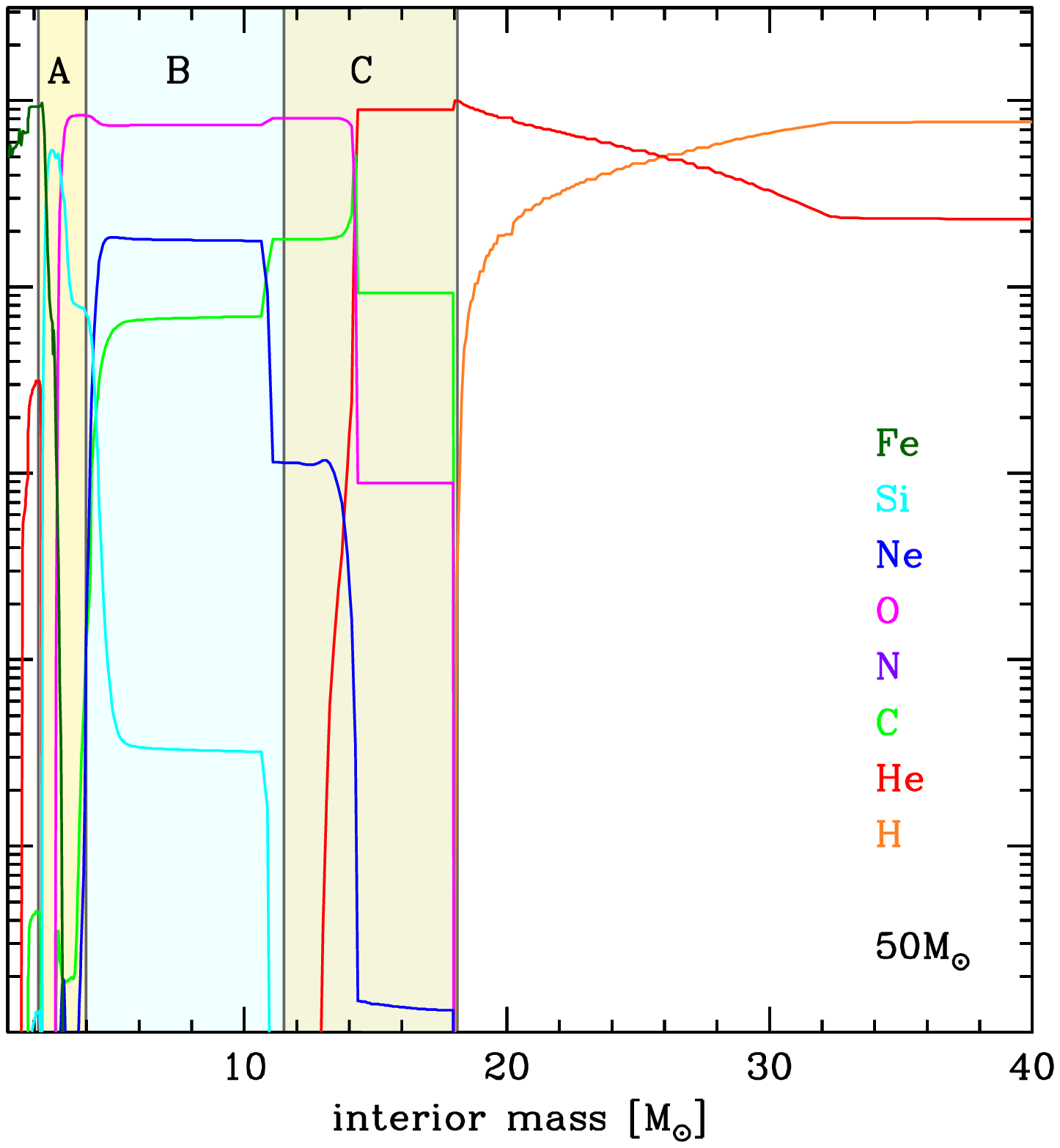}
\caption{Pre-supernova chemical structure for three selected models: 
$\rm 15 \, M_{\odot}$ (left panel), $\rm 30 \, M_{\odot}$ (central panel), $50 \, M_{\odot}$ (right panel). 
The shaded region extends up to the mass coordinate of the He core. Different shaded regions
indicate the layers that we will consider in the unmixed SN models (layers A, B, and C from
left to right). Colour versions of the figures are available online.}
\label{prog_diff}
\end{figure*}

\section{PopIII SN progenitors models}
\begin{figure*}
\includegraphics[width=7.50cm]{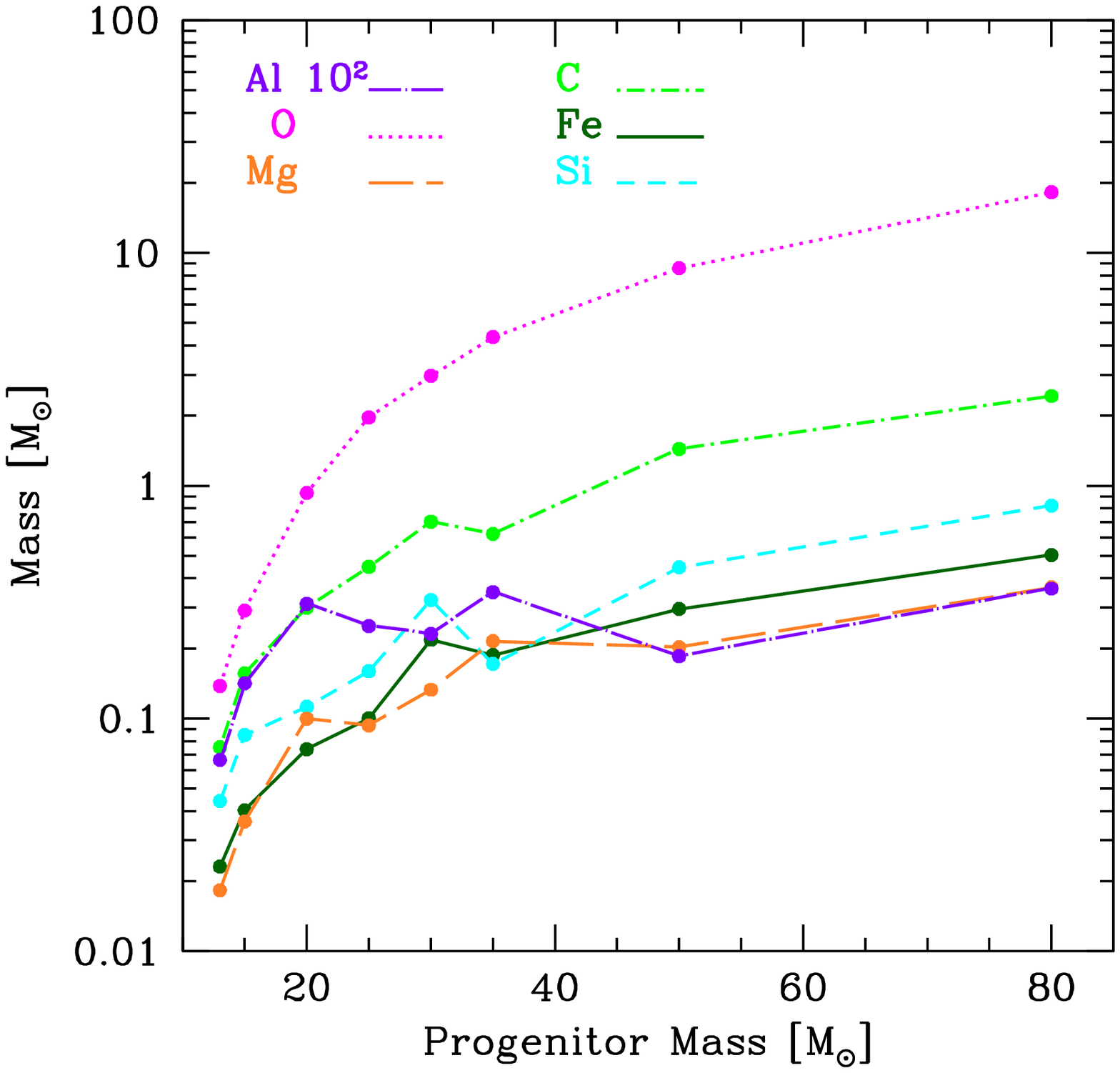}
\includegraphics[width=7.50cm]{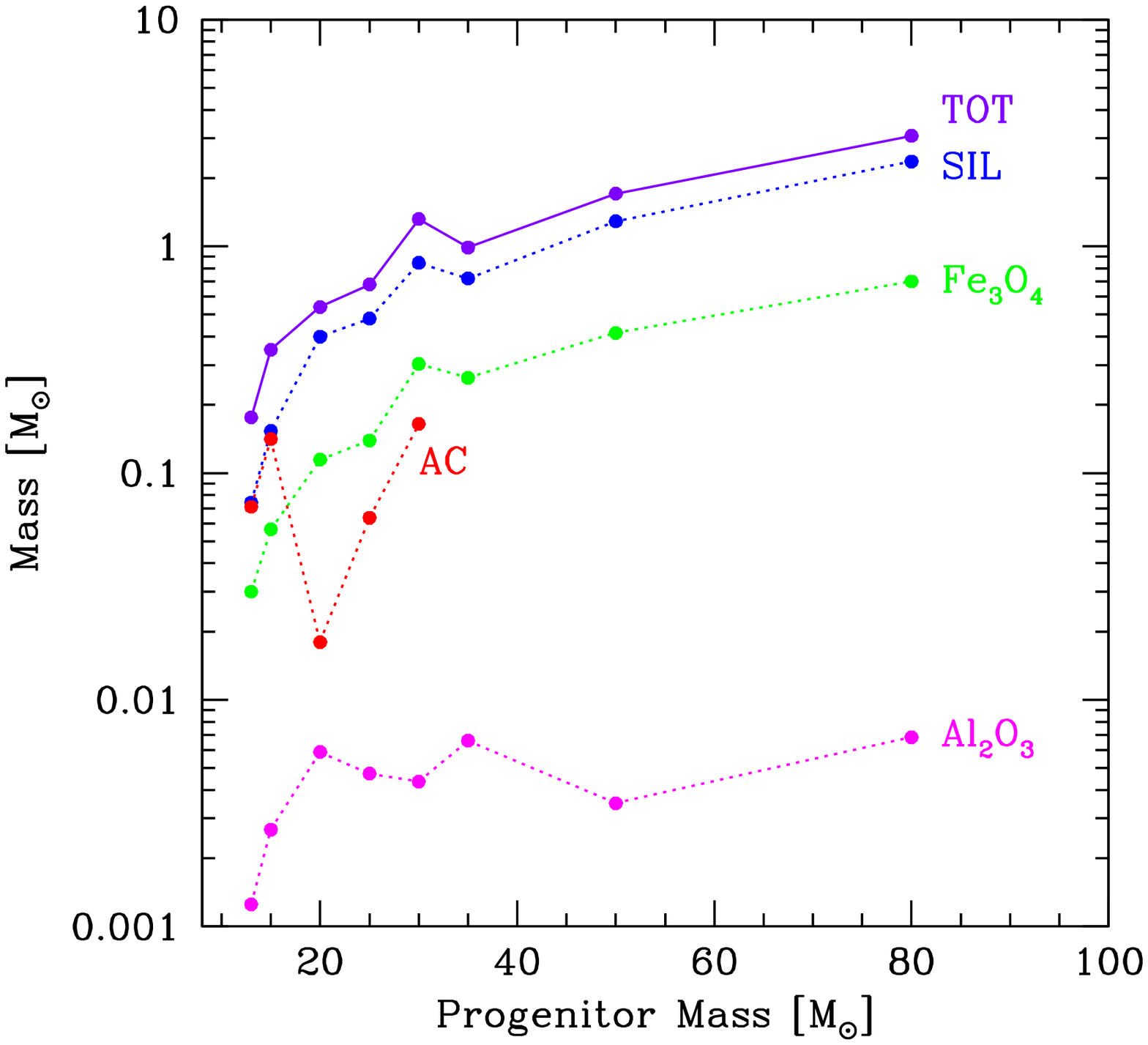}
\caption{Left panel: initial metal abundances in the ejecta of the adopted SN models as a function of 
the progenitor stellar mass (the abundance of Al has been multiplied by $10^2$). Right panel: mass of dust grains, before the passage of the reverse shock, 
as a function of the progenitor mass. $\rm SIL$ is the total mass in silicates, including \Forsterite, \Enstatite~and \Silica.}
\label{MpMg_MpMet}
\end{figure*}
\begin{figure*}
\includegraphics[width=8.0cm]{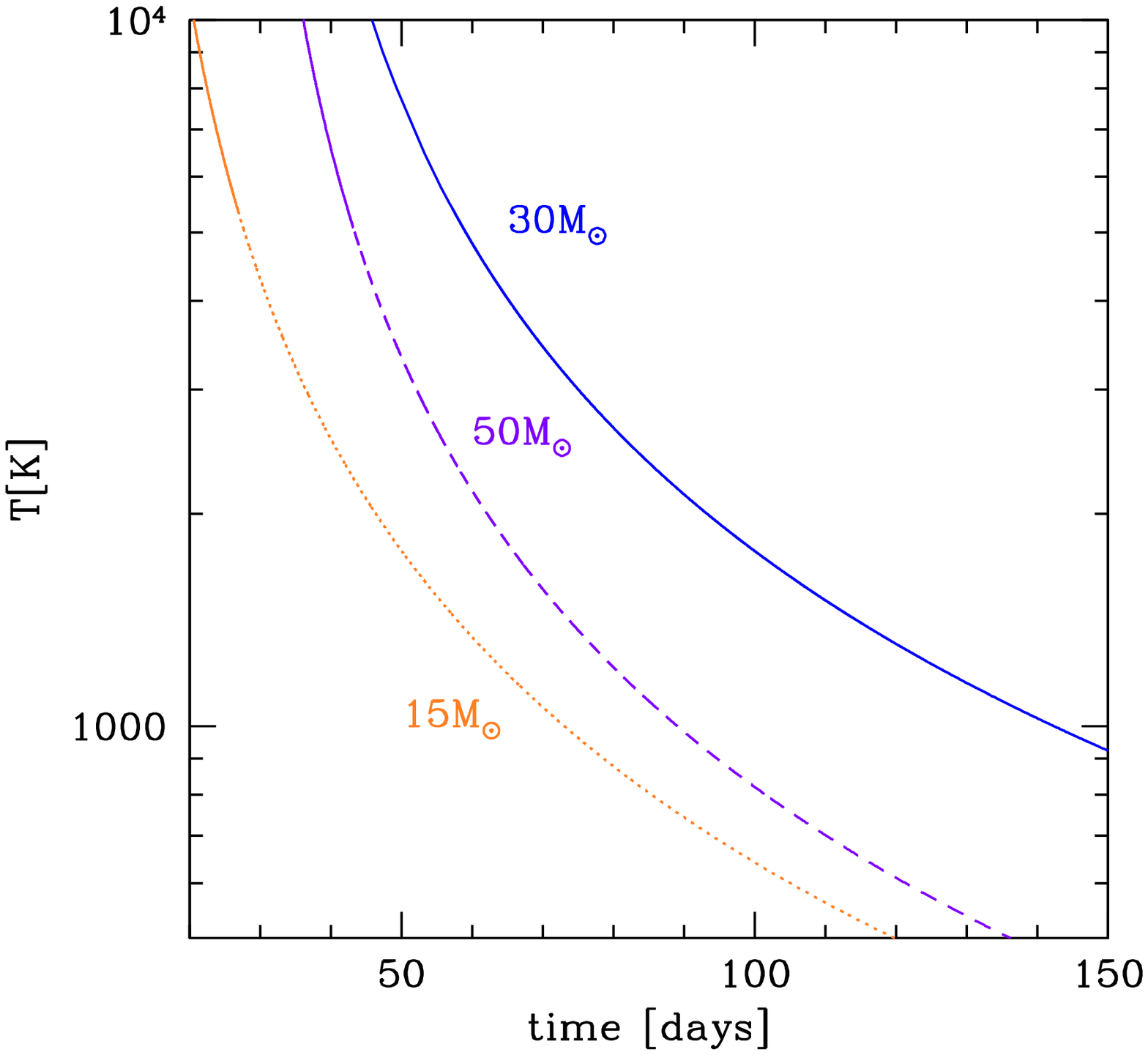}
\includegraphics[width=8.0cm]{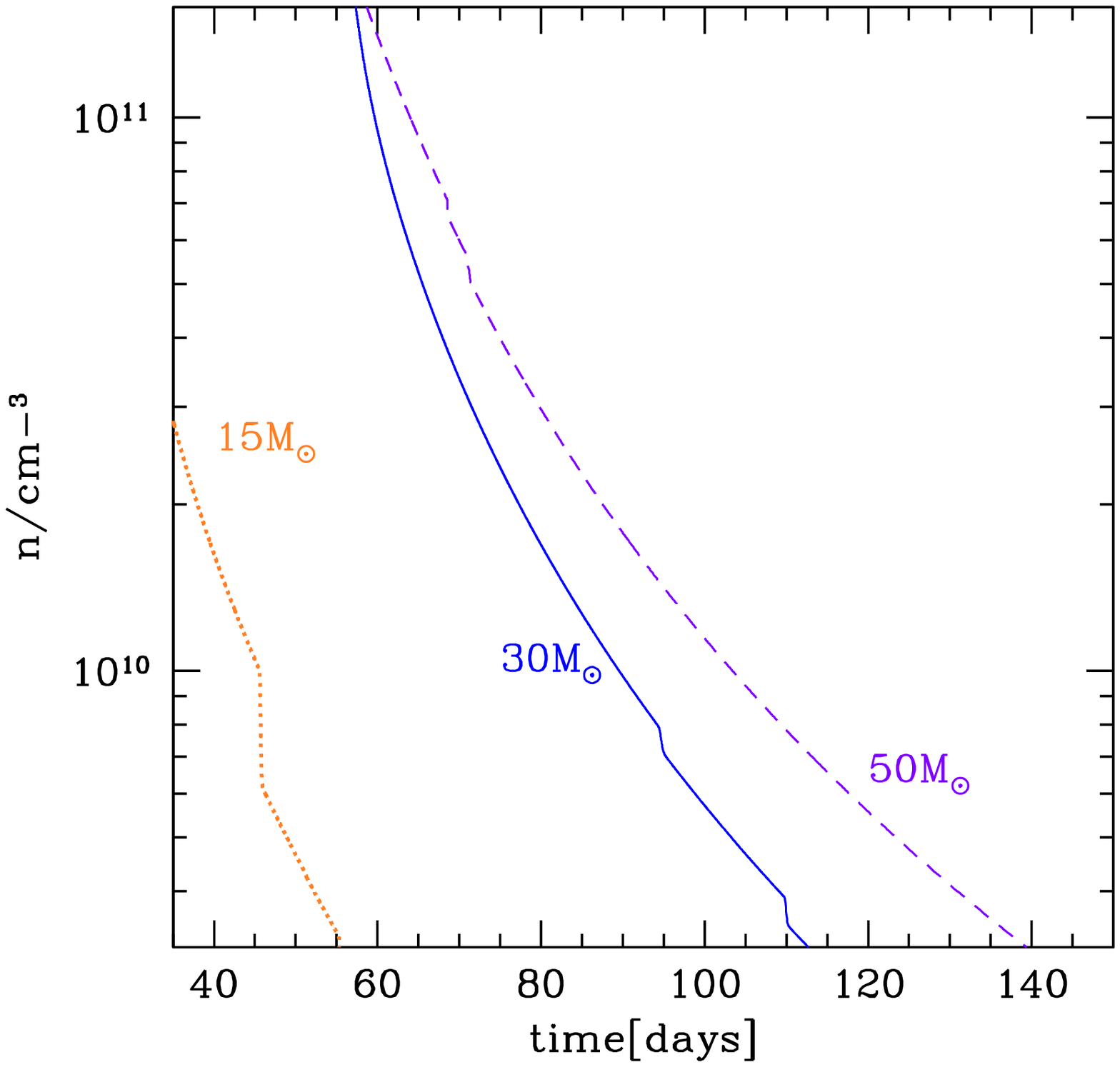}
\caption{Temperature (left panel) and number density (right panel) evolution for $15 \, M_{\odot}$ (solid), 
$30 \, M_{\odot}$ (dotted), $50 \, M_{\odot}$ (dashed) ejecta models.}
\label{Temp_dens_vs_time}
\end{figure*}
The presupernova models adopted in this paper are the ones presented and discussed 
in details in Limongi \& Chieffi~(2012). These models span a range of mass between 13 and 
80 \, M$_{\odot}$ and have a pristine Big Bang initial composition. The evolution of these 
models has been followed from the pre-main sequence up to the onset of the iron core 
collapse by means of the FRANEC stellar evolutionary code (Limongi \& Chieffi~2006). 
The explosion of the mantle and the consequent explosive nucleosynthesis have been 
computed in the framework of the "artificially induced explosion". Then, for each model, 
the mass cut ($\rm M_{cut}$), i.e. the mass coordinate which separates the final remnant
from the ejected portion of the mantle, has been fixed by requiring a best fit to the
element abundance pattern of the Cayrel average star, as extensively described in
Limongi \& Chieffi~(2012). In Table~\ref{tabappI:masses_normal} we summarize the main
properties of the SN explosion models. 

We construct the Pop~III ejecta models, requiring that the thermal, dynamical and chemical evolution 
of the ejecta evolve consistently with the explosive nucleosynthesis simulation 
(see next section for details). We have performed dust formation calculations 
assuming uniform mixing of the elemental abundances in the He core. However, due 
to the uncertainties related to the efficiency of mixing in metal-free SNe 
(Joggerst et al.~2009, Heger \& Woosley~2010), for some models we also present 
the results obtained assuming unmixed ejecta. In Fig.~\ref{prog_diff} we show the 
chemical structure (elemental mass fraction) as a function of the mass 
coordinate for three selected pre-supernova models, $\rm 15 \, M_{\odot}, 
30 \, M_{\odot}, and \, 50 M_{\odot}$ (for a detailed description of the differences 
emerging in the convective zones we refer the reader to Limongi \& Chieffi~2012). 
The shaded region extends up to the mass-coordinate of the He core, 
different colours indicate the layers that we will consider in the 
unmixed SN models, in Section 4. The left boundary of the innermost 
shaded region corresponds to the adopted mass-cut coordinate. In the 
left panel of Fig.~\ref{MpMg_MpMet} we show, for each SN model, 
the initial abundance of metal species which participate to molecules 
and dust formation; for all the SN models, the ejecta are very rich in 
carbon and oxygen and the total mass of metals is an increasing function
of the progenitor mass. The ejecta mass $\rm M_{eje}$, as expected, is an increasing 
function of the progenitor mass and explosion energy (see Table \ref{tabappI:masses_normal}).

\section{Results: PopIII SN dust yields}
\begin{figure*}
\hspace{-1.0cm}
\includegraphics[width=7.65cm]{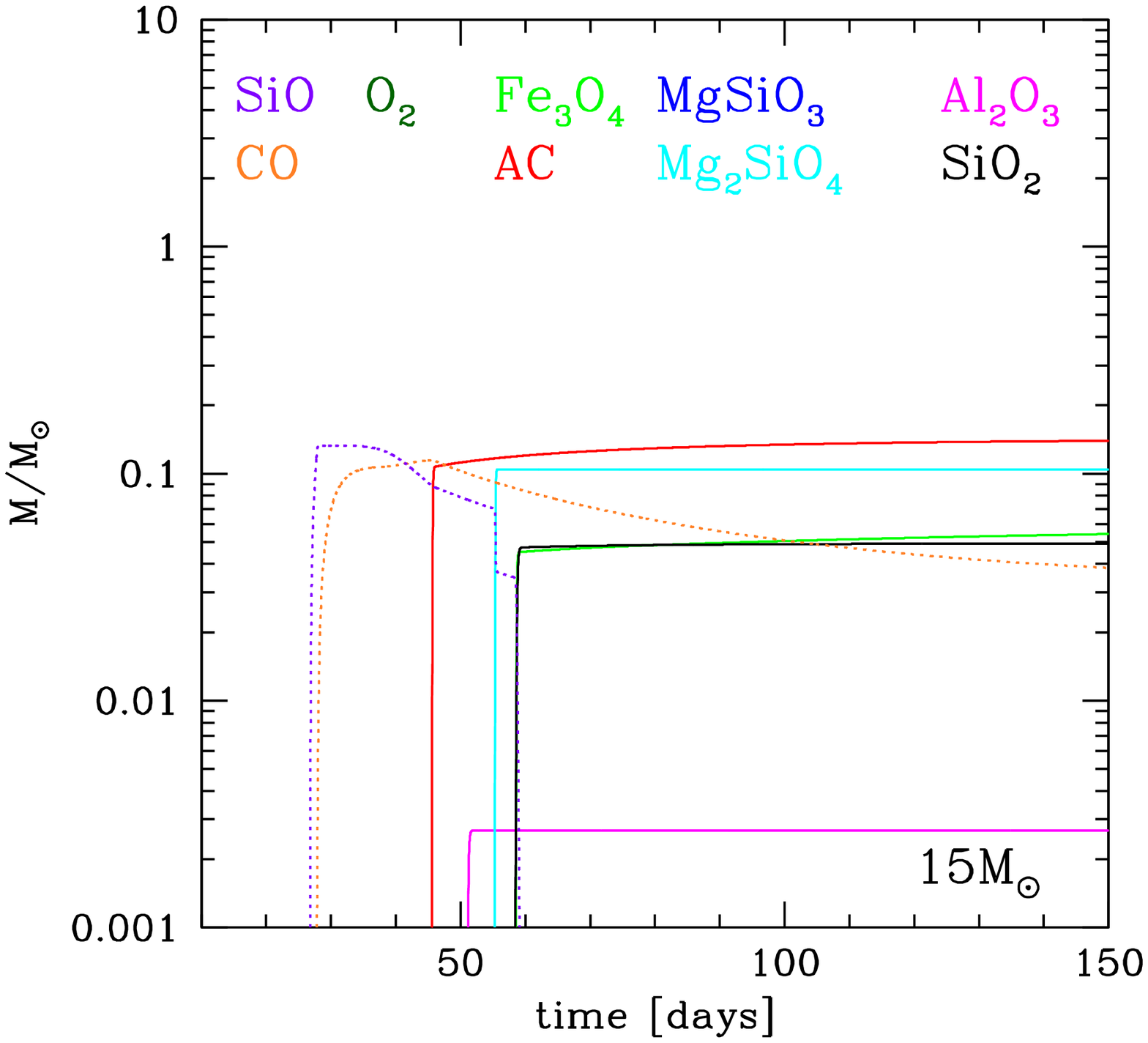}
\hspace{-2.5cm}
\includegraphics[width=7.65cm]{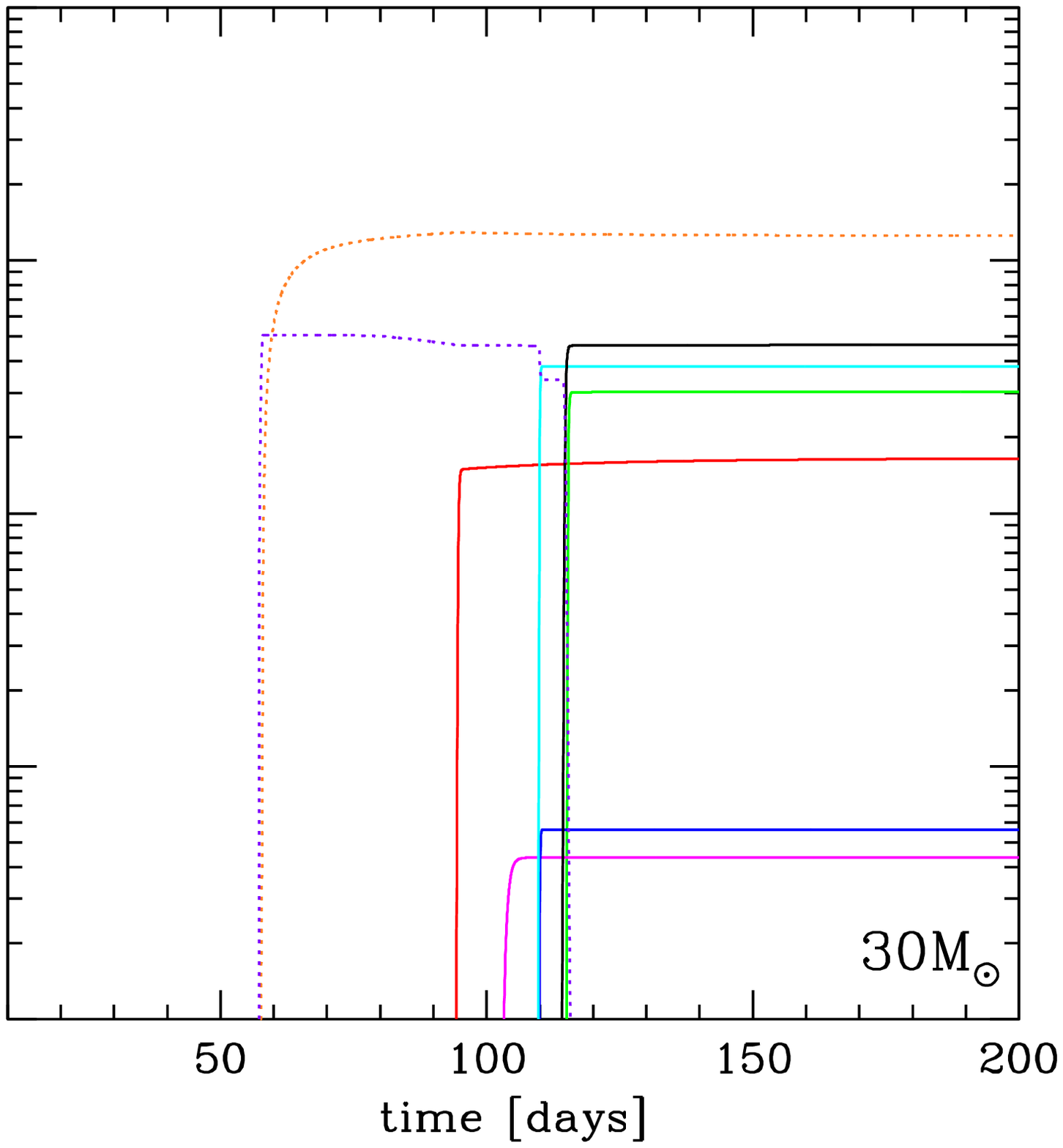}
\hspace{-2.5cm}
\includegraphics[width=7.65cm]{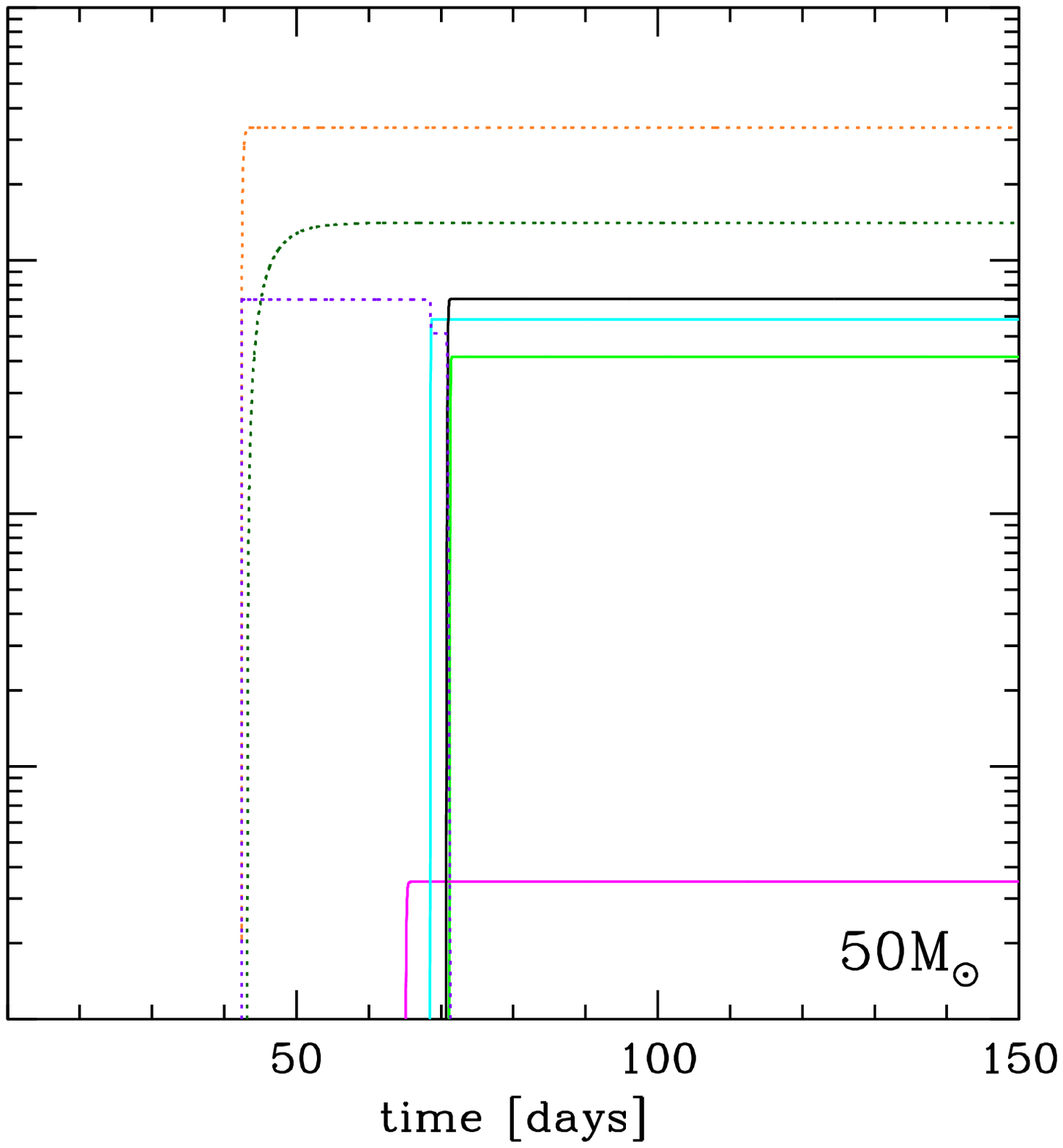}
\caption{Time evolution of molecular (dashed lines) and dust masses (solid lines) for three metal-free 
SN progenitors: 15 \,M$_{\odot}$ (left panel), 30\, M$_{\odot}$ (middle panel), 
50 \, M$_{\odot}$ (right panel).}
\label{MpMdustMmol}
\end{figure*}
This section presents the calculated dust yields that we obtain for Pop~III 
core-collapse SN models. As stated in the previous section, we construct 
the ejecta models using as initial conditions the thermo-dynamical properties 
obtained by the SN explosion simulation outputs (Limongi \& Chieffi~2012). 
The ejecta follow an adiabatic expansion and the temperature evolution is 
given by, 
\begin{equation}
\rm{T}=\rm{T_{\rm 0}}\left[1+\frac{\rm v_{eje}}{R_{\rm 0}}(t-t_{ini})\right]^{3(1-\gamma)}\quad \text{where}\quad 
\rm v_{eje}=\sqrt{\frac{10 \rm E_{\rm expl}}{3\rm M_{\rm eje}}}\,
\label{thermo}
\end{equation}  
is the ejecta expansion velocity, $\gamma=1.41$ is the 
adiabatic index, and $\rm T_{0}$ and $\rm R_0$ are the temperature and radius 
of the He core at the initial time $\rm t=t_{\rm ini}$. This initial time $\rm t_{ini}$ 
is fixed by requiring that the gas temperature at the radius of the He core, $\rm R_{He_{core}}$ reaches a 
temperature of $\rm T_{0}=10^{4}$\,K. 
For all uniformly mixed SN ejecta models (labelled with the progenitor mass)
Table~\ref{tabappI:masses_normal} reports the thermodynamical properties, 
the metal yields of the key elements in the nucleation process, the total 
amount of dust, the mass of molecules and the mass of dust in each grain 
species. In what follows, we discuss these results in details.

\subsection*{Dust formation in uniformly mixed ejecta}
\begin{figure*}
\hspace{-1.0cm}
\includegraphics[width=7.65cm]{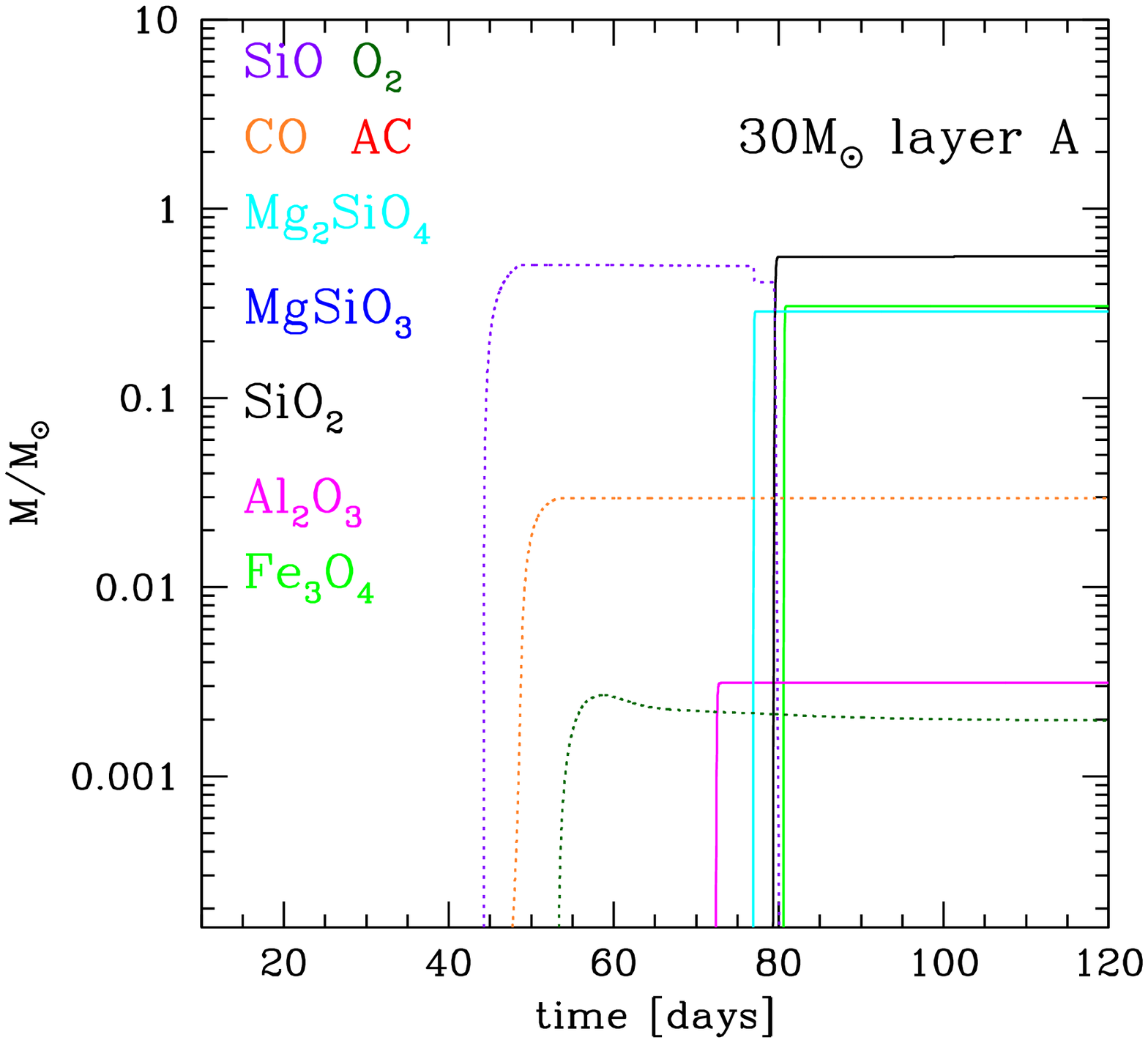}
\hspace{-2.5cm}
\includegraphics[width=7.65cm]{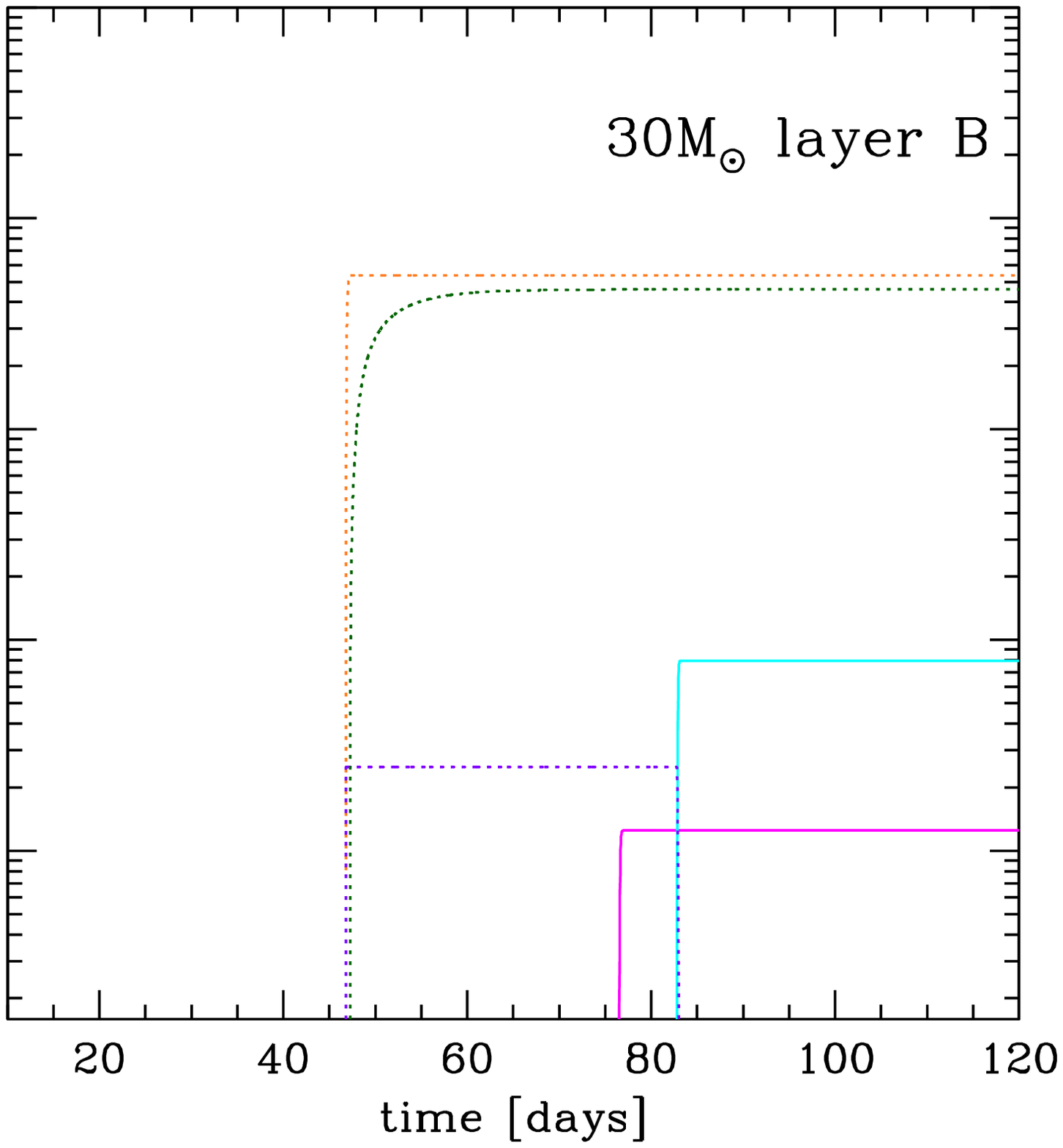}
\hspace{-2.5cm}
\includegraphics[width=7.65cm]{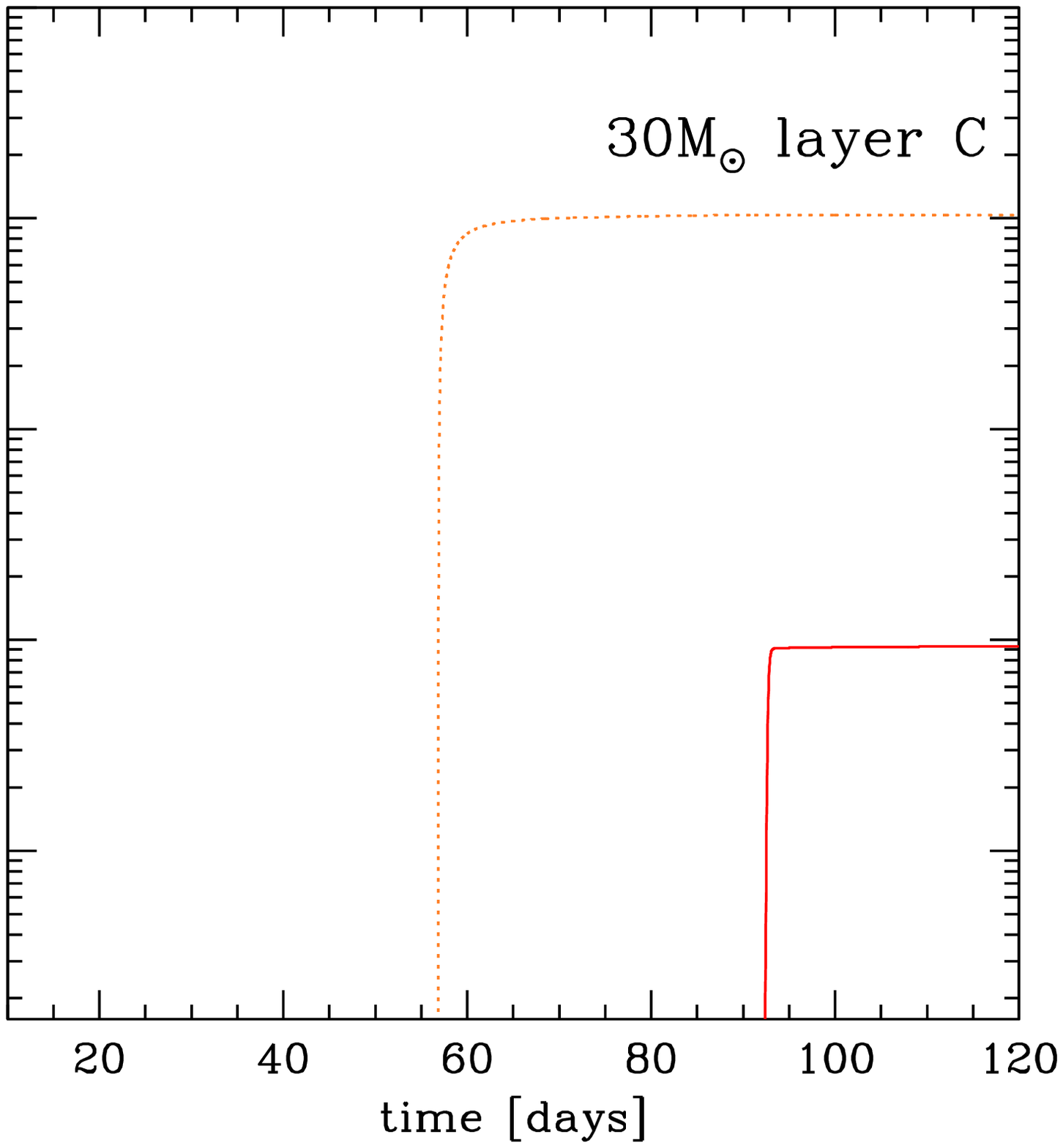}
\caption{Time evolution of molecular (dashed lines) and dust (solid lines) masses for the $\rm 30 \,M_{\odot}$ unmixed model: 
layer A (left panel),  layer B (middle panel), 
layer C (right panel). Colour versions of the figures are available online.}
\label{MpMdustMmol_unmixed}
\end{figure*}
 
In the right panel of Fig.~\ref{MpMg_MpMet} we plot the mass of dust
for different grain species as a function of the progenitor mass. We find that: 
(i) the total dust mass produced increases with the progenitor mass, ranging between 
$[0.18 - 3.1]$~M$_{\odot}$, and it is dominated by silicates; 
(ii) the second most abundant dust species is magnetite, due to the initially 
high iron abundance present in the ejecta (see left panel of Fig.~\ref{MpMg_MpMet}) 
plus the iron produced by $\rm ^{56}Ni$ decay; (iii) amorphous carbon forms 
only in less massive models, with progenitor masses $\rm \le 30\,M_{\odot}$. This is due to the 
larger mean ejecta density of more massive progenitors, 
which increases the rates of the major processes leading to 
the formation of CO molecules, locking carbon atoms and 
preventing the formation of AC grains (we will return to this 
point later); (iv) alumina is the less abundant dust species, reflecting the lower 
abundance of Al in the ejecta. 

Using three representative SN progenitors with masses $\rm 15 \, 30$ and $\rm 50 \, M_{\odot}$, 
Fig.~\ref{Temp_dens_vs_time} shows the evolution of the ejecta temperature and number density: 
when the He core of the three SN progenitors reaches $\rm T_{0}=10^{4}$K, we fixed the initial 
time $\rm t_{ini}$, which varies between (1.78 - 3.96)$\times 10^{6}$s 
(see Table~\ref{tabappI:masses_normal}). At these initial times, the three ejecta models have
 radii of $2.44\times 10^{14}$(15 M$_{\odot}$), $4.68\times 10^{14}$(30 M$_{\odot}$), and
$2.51\times 10^{14}$(50 M$_{\odot}$) cm, and start the adiabatic expansion with velocities in the
range (2960-3086)km s$^{-1}$ (see eq.~\ref{thermo}), and the temperature decreases.
As shown in the right panel of Fig.~\ref{Temp_dens_vs_time}, the initial number densities, 
$\rm n_0=$ 7.18$\times 10^{11}$(15 M$_{\odot}$), 
$7.39\times 10^{11}$(30 M$_{\odot}$), $1.24\times 10^{13}$(50 M$_{\odot}$) cm$^{-3}$, also decreases, 
showing a step-decrease corresponding to the condensation of the gas-phase elements.

In standard nucleation theory, dust grains condense when the gas becomes supersaturated.
For each grain species, this depends on the temperature, density and abundances of the
corresponding key element in the ejecta. Fig.~\ref{MpMdustMmol} shows that grains form 
between 45 and 120 days after the explosions. Although these timescales are smaller than 
the values found by Nozawa et al. (2003) for core-collapse Pop III SN with a 25 $\rm M_\odot$
progenitor, due to the different thermal evolution of the ejecta, the resulting dust masses and grain
size distribution are consistent because grain condensation occurs in similar physical conditions.

Molecules form according to the formation/destruction rates described in Section 2.1 and 
are also partially destroyed by the interaction with Compton 
electrons produced by destruction reactions rates $\rm D_{1}-D_{4}$ (see Table~2), 
up to the point where the temperature reaches the condensation
window and dust grains start to form. The ejecta density is a strong function
of the initial progenitor mass: when the ejecta temperature enters 
the condensation regime, at around T $\sim 2000$~K, the corresponding density
can vary by one order of magnitude, although the dependence on the progenitor mass
is not monotonic. Clearly, the  chemical evolution of the ejecta depends 
on the temperature and density at each given time: in Fig.~\ref{MpMdustMmol} 
we show, for the same set of SN models, the time dependence of the mass of CO, SiO and of the newly formed 
dust species.

We start analizing the $\rm 15 \, M_{\odot}$ model plotted in the left panel of 
Fig.~\ref{MpMdustMmol}: CO and SiO start forming at around 27 days after the
explosion, when $\rm T \sim 5400\,K$. At this time, the CO formation rate is dominated by the 
neutral process NN2 that rapidly depletes all the available $\rm C_{2}$. 
SiO molecules form through the neutral channel NN5 and radiative association 
process RA2. At around 44 days, when $\rm T \sim 2000\,K$, the impact with 
Compton electrons ($\rm D_{1}$) dominates over the other rates and partially destroys  CO molecules. 
The free carbon atoms form AC, which is the first 
dust species to form, due to its higher condensation temperature. 
When the temperature decreases to $\sim \rm 1764 \,K$, alumina grains form, which 
rapidly deplete aluminum from the ejecta. When T drops to $\rm \sim 1530$\,K,  \Forsterite~starts 
forming - at the same time of \Enstatite - but \Forsterite  reaches supersaturation first 
(because it has the largest nucleation current), rapidly consuming Mg and SiO and inhibiting the 
growth of \Enstatite. Finally, when $\rm T \sim 1400$\,K, \Magnetite \, and \Silica \, forms, depleting 
completely the remaining SiO and iron from the ejecta.

The central panel of Fig.~\ref{MpMdustMmol} shows the evolution of molecules and dust grains for 
the $\rm 30 \,M_{\odot}$ model: at $\sim 57$ days, when $\rm 5400 \leq T \leq 4250$, the dominant process is 
radiative association of $\rm C_{2}$ which - due to the large oxygen abundance -
is then rapidly depleted in CO through the reaction $\rm O+C_{2}\rightarrow CO+C$. 
When T decreases, at around 66 days, 
the dominant CO formation process becomes radiative association RA1. 
Around 93 days, the temperature reaches $\rm \sim 2000 \,K$, nucleation starts, D1 efficiently destroys 
the CO molecule, leaving C atoms free to form AC and - as in the $\rm 15 \,M_{\odot}$ model - the CO formation 
processes are not equally efficient in re-forming the molecule. As a result, AC grains form.

For the more massive model the evolution changes, as shown in the right panel of 
Fig.~\ref{MpMdustMmol}. The interplay between the larger ejecta 
density and the richer oxygen content (the mass of oxygen is $18\%$ of the mass of the ejecta) 
leads to the efficient formation of $\rm O_{2}$ molecules through radiative association RA4 and 
also activates efficiently the bimolecular process NN1, which exceeds the destruction rate D1, and 
becomes the leading process in CO formation. This depletes all the C atoms, inhibiting AC 
grains condensation. As for the two previous models, at $\rm 5400\leq T \leq 5000$ the 
main CO formation channel is $\rm O+C_{2}\rightarrow CO+C$. When T decreases, it 
becomes $\rm C+O_{2}\rightarrow CO+O$ and it remains so for the subsequent evolution. 
At 64 days, the nucleation process starts, the grain condensation proceeeds as 
described for the two previous models, because the condensation sequence of
the grain species reflects the corresponding condensation temperatures (Todini \& Ferrara 2001; 
Schneider et al. 2004). 
 
We summarize our results on mixed ejecta models in Table~\ref{tabappI:masses_normal}. 
In general, the total mass of CO molecules ranges between $\rm [9.60 \times 10^{-3} - 5.66]\,M_{\odot}$,
increasing with the mass of the ejecta due to the initial abundances of oxygen and carbon. 
SiO molecules form efficiently 
through NN5 (at early times) and through RA2, but it is completely depleted 
in silicates at the end of the nucleation process. $\rm C_{2}$ molecules 
are consumed at very early time in the formation process of CO, through 
the channel NN2. For models less massive than $\rm 30 \, M_{\odot}$, 
$\rm O_{2}$ is under-produced (and consumed rapidly in CO formation 
process), with respect to more massive models. When $\rm M > 30 \, M_{\odot}$ 
 the higher mean ejecta density and oxygen abundance cause  radiative 
association rate RA4 to become the dominant process. This enables a very 
efficient $\rm NN1$ reaction rate which overcomes the destruction 
rate $\rm D_{1}$, locking all carbon atoms in CO molecules.

\subsection*{Dust formation in umixed ejecta}
The mixing efficiency in SN events is still debated, due to 
complexity of physical processes associated to core-collapse, 
such as asphericity and growth of Rayleigh-Taylor instabilities which accompany shock propagation. 
Besides, these effects are multi-dimensional and this adds an extra source 
of uncertainty related to emerging differences in 1D-3D numerical simulations 
(see Joggerst et al.~2009, Heger \& Woosley 2010 and references therein).
In particular, the occurence and the growth of Rayleigh-Taylor instabilities 
depends on stellar mass and metallicity.  It has been shown that the more 
compact is the star, the more rapid is the reverse shock propagation, giving less 
time to these hydrodynamical instabilities to grow (Joggerst et al.~2009). 
For this reason, we have decided to investigate the impact of ejecta mixing
on dust formation and we have considered stratified models, as the limiting
cases when Rayleigh-Taylor instabilities do not grow. The unmixed case is 
based on the hypothesis that, due to inefficient mixing, the elemental abundances 
reflects the pre-supernova stratified composition: this means that more internal 
elements, such as magnesium and silicon are not present or present in very little
amount in the more external layers of the star, reducing the probability to form silicates.
\begin{figure}
\includegraphics[width=8.50cm]{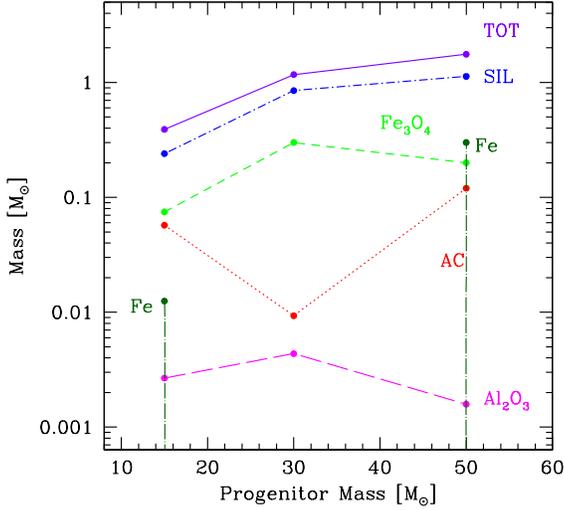} 
\caption{Mass of dust grains, before the passage of the reverse shock, 
as a function of the progenitor mass for the three selected unmixed models. 
$\rm SIL$ is the total mass in silicates, including \Forsterite, \Enstatite~and \Silica.}
\label{dust_unmixed}
\end{figure}

In Fig.~\ref{prog_diff}, for each progenitor model, the shaded regions 
represent the layers that we consider in the stratified models, which 
extend from the mass cut coordinate to the outer radius of the He core ($\rm R_{He_{core}}$).  
The number and extent of the layers have been selected based on the criterium
of constant abudances of the relevant elemental species within each layer. Hence,
the outer radius of layer A corresponds to the mass coordinate where the Si abudance
rapidly drops, and the outer radius of layer B is where the abundance of C (Ne) rapidly
increases (decreases).
 
For the $\rm 15, 30$ and $\rm 50\,M_{\odot}$ models, we report in 
Tables~\ref{tabappII:masses_unmixed_15},~\ref{tabappII:masses_unmixed_30} and
~\ref{tabappII:masses_unmixed_50} the thermodynamical properties, 
the initial elemental composition, the masses of molecules and dust 
for all the layers. The time evolution of the molecular and dust masses for each of the
three layers of the  $\rm 30 \, M_{\odot}$  progenitor is shown in Fig.~\ref{MpMdustMmol_unmixed}. 
In the stratified ejecta, due to the different temperature 
evolution of the layers, CO starts to form earlier in the evolution with respect 
to the corresponding mixed case, and the same is true for SiO, where present. 
It is worth to note that, compared to the mixed case where $\rm O_{2}$ 
is completely depleted in CO, in layers A and B the $\rm O_{2}$ molecule forms, 
while AC forms only in layer C. Silicates and alumina form in the two internal 
layers A and B where Al, Si and Mg are present. Magnetite forms only in layer 
A where iron is present. The total mass of dust for the $\rm 30 \, M_{\odot}$ progenitor 
is $\rm 1.2 \,M_{\odot}$, about $90\%$ of the mass formed in the mixed case. 

Fig.~\ref{dust_unmixed} summarizes the results obtained for unmixed ejecta models. The mass of dust for 
different grain species is shown as a function of the progenitor mass. We find that the 
total dust mass ranges between $[0.39 - 1.76]$~M$_{\odot}$, and it is dominated by silicates. 
Contrary to the fully mixed cases, for the $\rm 15 \,M_{\odot}$ and $\rm 50 \,M_{\odot}$  
SN progenitor models, solid iron is able to form due to the initially high iron abundance 
present in the most internal layer of the ejecta plus the iron produced by $\rm ^{56}Ni$ decay.
As a result, in these models the total dust mass formed is $\sim 3 - 10\%$ larger than in the mixed case.
The CO mass, differently from the mixed case, is not increasing with the 
progenitor mass, but varies depending on rate formation efficiency which 
is a function of temperature, mean density and composition of each 
layer.

\subsection*{Dust destruction by the reverse shock}
\begin{figure*}
\includegraphics[width=8.70cm]{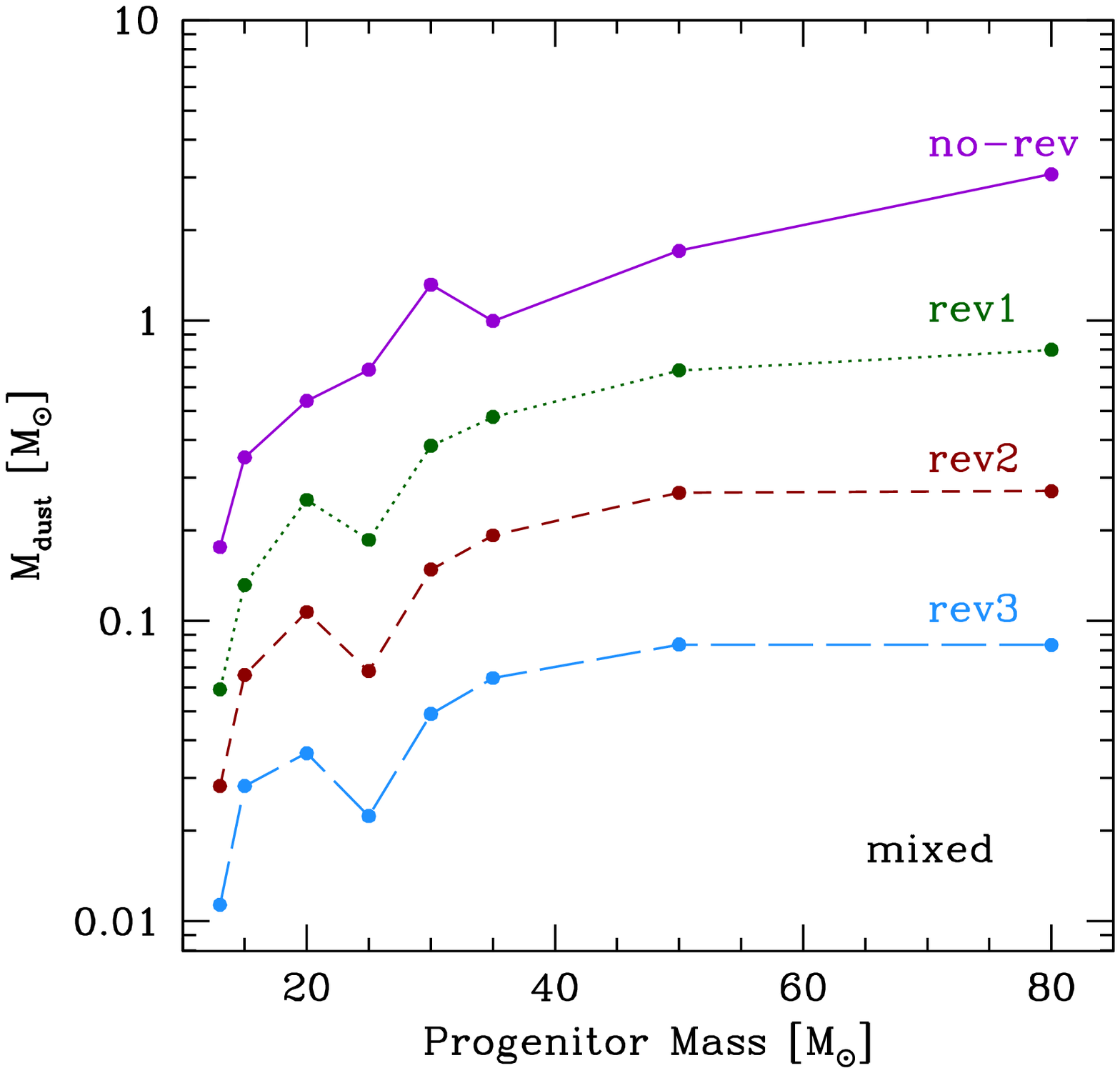}
\includegraphics[width=8.70cm]{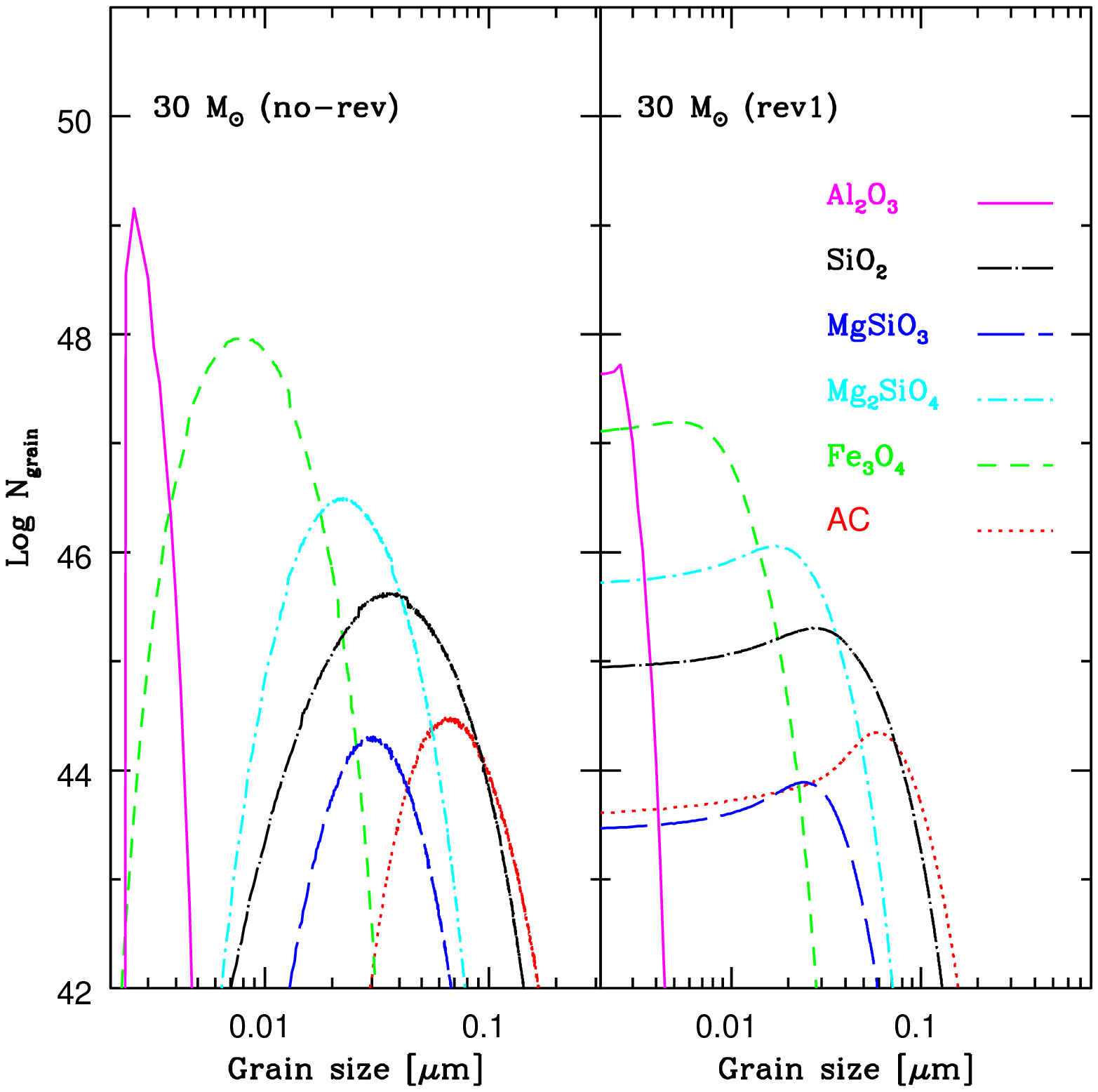}
\caption{Left panel: the mass of dust at the end of nucleation, 
and after the passage of a reverse shock of increasing intensity for all the mixed SN models 
considered in the present study. From top to bottom: no reverse shock 
models ({\bf no-rev}), models with a circumstellar medium density of 
$\rm n_{ISM} =0.06$ ({\bf rev1}), $0.6$,({\bf rev2}), and $6 \,\rm cm^{-3}$ ({\bf rev3}).
Right panel: size distribution function of the grains for a  
$\rm 30 \, M_{\odot}$ progenitor model  before ({\bf no-rev}, left) and after the passage of the
reverse shock ({\bf rev1}, right).}
\label{fig:revshock}
\end{figure*}
To describe the impact of the reverse shock on the dust grains 
formed inside the expanding ejecta and to estimate the surviving 
dust mass, we follow the approach described in Bianchi \& Schneider 
(2007) for which we summarize the key points: (i) the dynamics of 
the reverse shock is treated using Truelove \& McKee analytic 
approximations (Truelove \& Mckee 1999) which follows the forward 
and reverse shock evolution as a function of the main ejecta 
parameters, such as kinetic energy, $\rm E_{expl}$, ejecta mass, 
$\rm M_{eje}$, and density of the ISM, $\rm \rho_{ISM}$; 
(ii) the distribution of dust grains in the ejecta is uniform 
and the size distribution is the same everywhere; (iii) to 
quantify the role of $\rm \rho_{ISM}$ we analize three different 
cases with $\rm n_{ISM}= 0.06 ({\bf rev1}), 0.6 ({\bf rev2}), and \, 
6 cm^{-3} ({\bf rev3})$. 
As previously shown in Todini \& Ferrara~(2001) and Nozawa \& 
Kozasa~(2003), the nucleation and accretion processes lead to a 
typical lognormal grain size distribution: in the right panel of Fig.~\ref{fig:revshock} 
we show the grain size distribution of the $\rm 30 \,M_{\odot}$ SN model
with mixed ejecta, before and after the passage of the reverse shock. 
The grain sizes range between
$\rm [10^{-3} - 0.5]\mu m$, depending on the grain species and on the
SN ejecta models. Since AC is 
the first grain species to condense, it has sizes larger than the 
other grains because grain accretion is more efficient at larger densities. 
For the same reason, alumina grains have the smallest sizes due to the 
low initial abundance of Al in the ejecta. As a result, the latter grain species
is almost completely destroyed by the reverse shock, while the other grain
species suffer a partial destruction, with the grain size distribution flattening
towards smaller grain sizes.

\begin{figure*}
\hspace{-1.4cm}
\includegraphics[width=7.0cm]{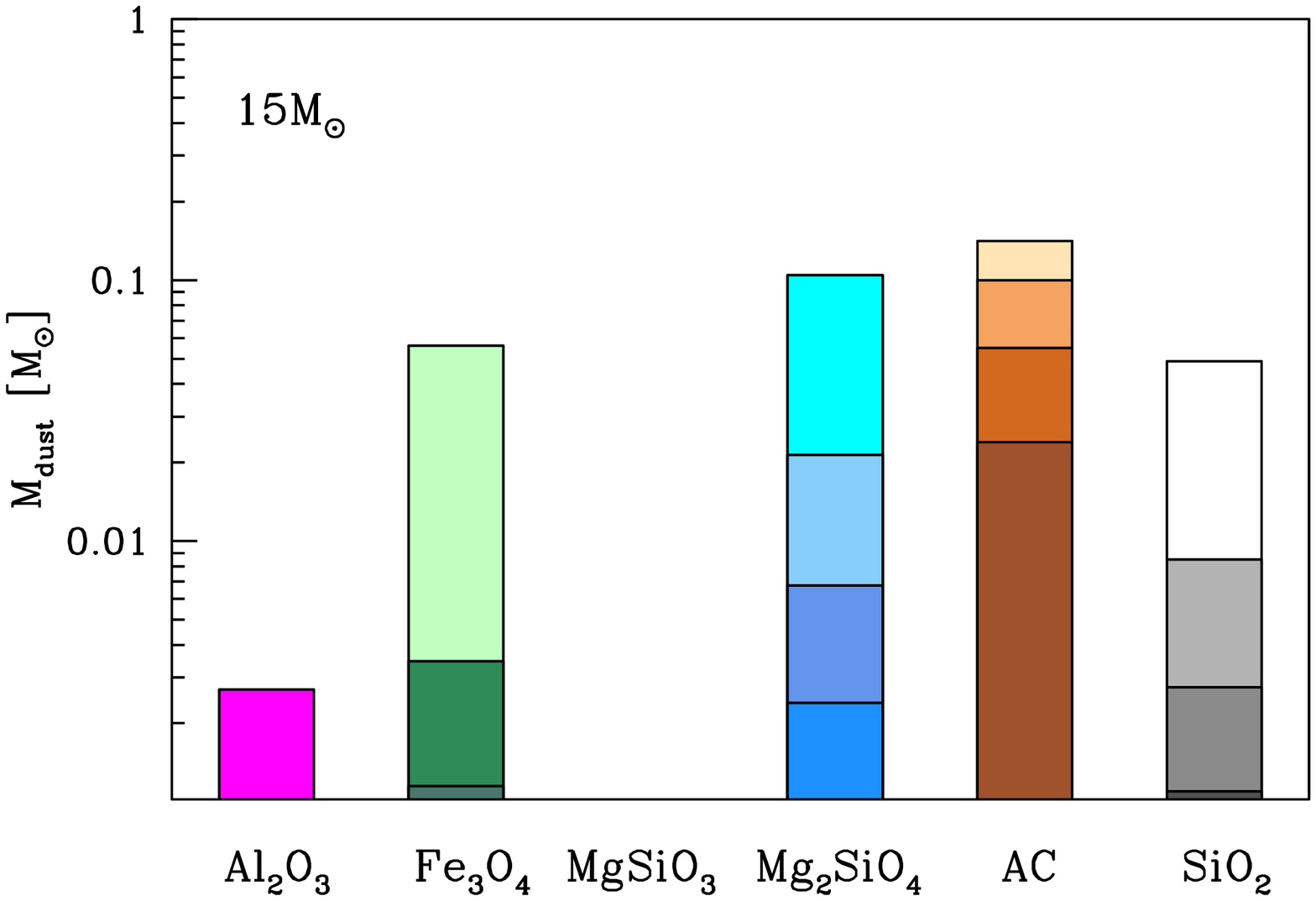}
\hspace{-1.45cm}
\includegraphics[width=7.0cm]{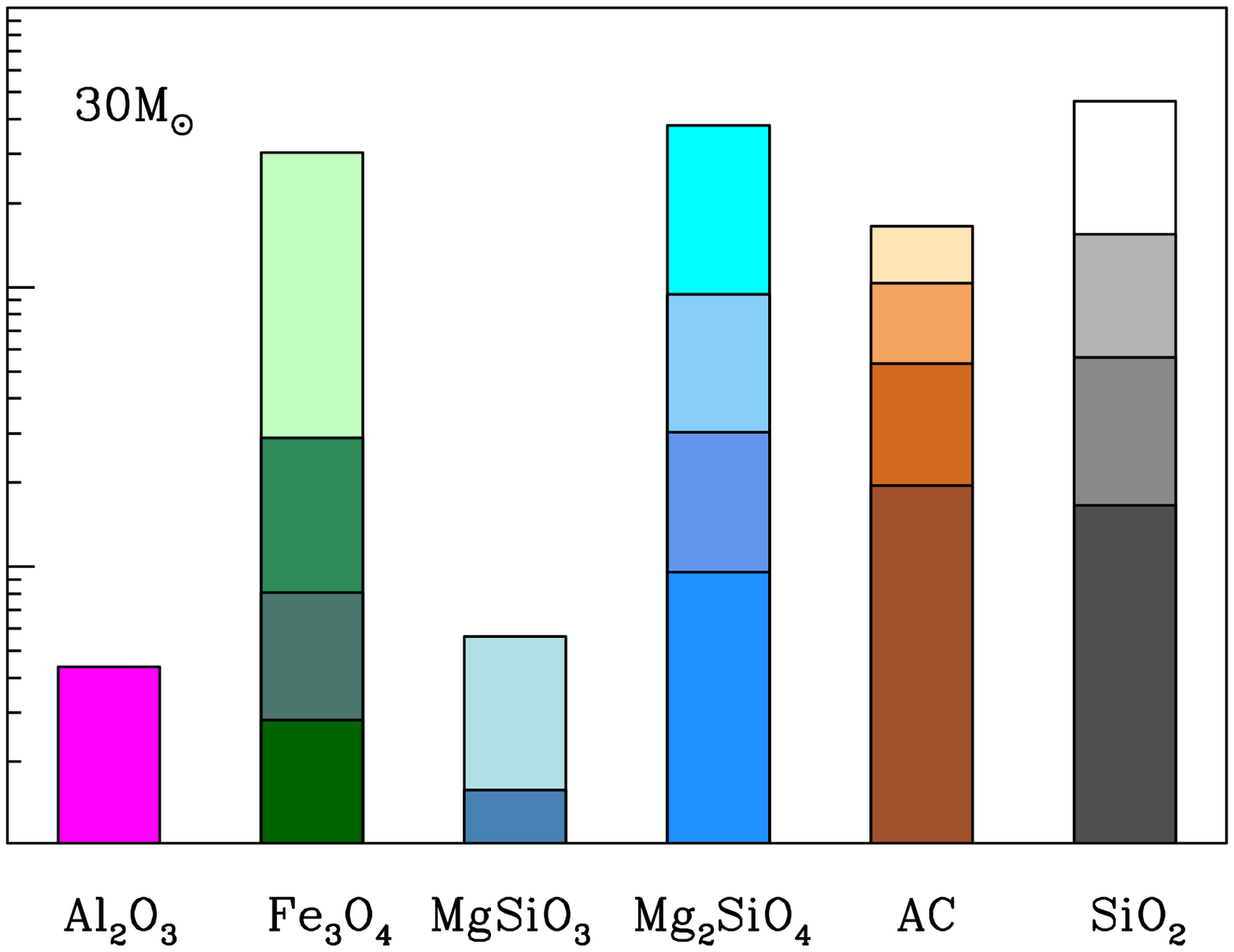}
\hspace{-1.45cm}
\includegraphics[width=7.0cm]{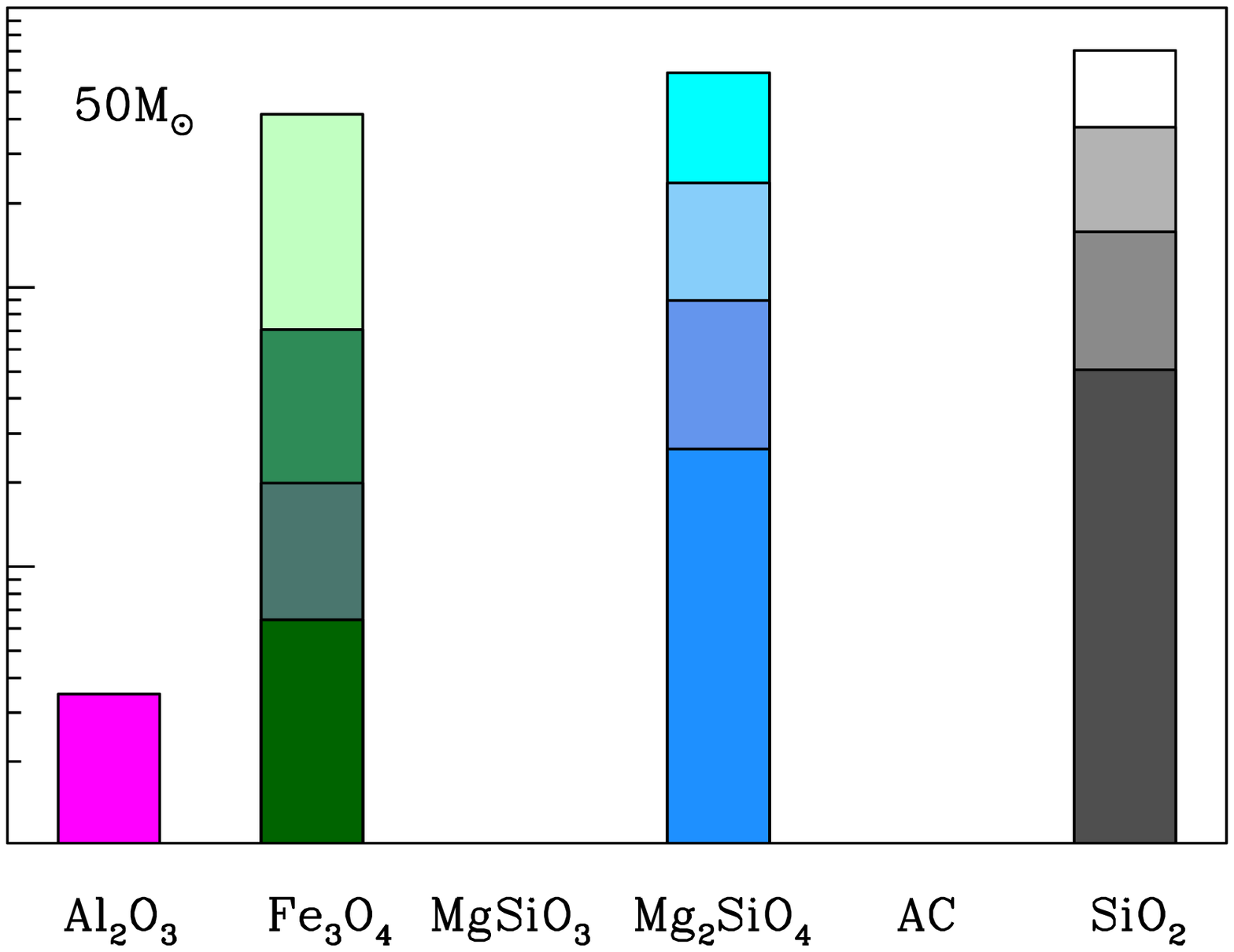}
\hspace{-1.45cm}
\caption{Histograms showing the mass of dust in the different compounds at the end of nucleation, 
and after the passage of a reverse shock of increasing intensity for three mixed SN models of 15M$_{\odot}$, 30M$_{\odot}$ 
and 50M$_{\odot}$. From top to bottom: no reverse shock models ({\bf no-rev}), models with a circumstellar medium 
density of $\rm n_{ISM} = 0.06 \, ({\bf rev1}), 0.6 \,({\bf rev2}) and \, 6 \, \rm  cm^{-3}$ 
({\bf rev3}).}
\label{3revmodels}
\end{figure*}

The dust mass which survives the impact of the reverse shock 
is reported for all  mixed SN models in Table~\ref{rev_core_collapse}, 
where we have also specified the dust mass of each grain species. 
In the left panel of Fig.~\ref{fig:revshock} we show, for all Pop~III mixed SN models, 
from top to bottom, the total mass of dust at the end of nucleation 
and after the impact of the reverse shock for the three increasing values of $\rm n_{ISM}$. 
In the worst scenario, when the SNe explode in the densest ISM, 
the reverse shock travels faster and encounters a higher density 
gas inside the ejecta, increasing the sputtering and causing that 
only $\sim 3\%$ of the newly formed dust mass survives. Clearly, the percentage of 
surviving dust depends on the ejecta model and ranges from $48\%$ 
to $3\%$. In Fig.~\ref{3revmodels} the same histograms are reproduced 
for $\rm 15, 30$ and $\rm 50 \,M_{\odot}$ mixed models, showing 
the impact of the reverse shock on the different grain species. 
We can see that alumina grains are completely destroyed. 
Depending on $\rm n_{ISM}$, after the passage of the reverse 
shock, we see that: (i) in the $\rm 15 \, M_{\odot}$ model the dominant 
species becomes AC, differently from the {\bf no-rev} 
case, where AC and silicates have comparable masses; (ii) in the 
$30 \, M_{\odot}$ model silicates always dominate the mass of dust;
however, the smaller destruction suffered by AC grains, 
compared to other grain species, alters the original dust composition 
and after the passage of the reverse shock the mass of AC and silicates 
are comparable and AC grains are more abundant than $\rm Fe_{3}O_{4}$ grains; 
(iii) in the $\rm 50 \,M_{\odot}$ model,  silicates dominate 
all the grain species, as in the {\bf no-rev} case. 

\begin{figure}
\includegraphics[width=8.50cm]{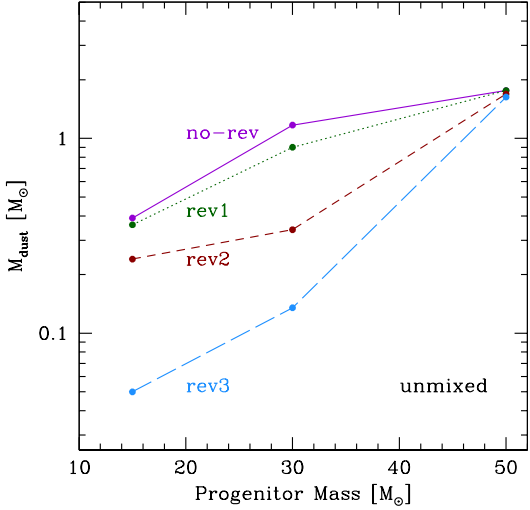} 
\caption{Mass of dust grains, after the passage of the reverse shock, 
as a function of the progenitor mass for the three selected unmixed models. 
From top to bottom: no reverse shock models ({\bf no-rev}), models with a circumstellar medium 
density of $\rm n_{ISM} = 0.06 \, ({\bf rev1}), 0.6 \,({\bf rev2}) and \, 6 \, \rm  cm^{-3}$ 
({\bf rev3}).}
\label{dust_unmixed_rev}
\end{figure}

Fig.~\ref{dust_unmixed_rev} illustrates the dust mass which survives 
to the impact of the reverse shock for the three selected unmixed models 
(see also Table \ref{rev_unmixed} for the amount of grains of different species in the 
three layers).
After the passage of the reverse shock, depending on the density of the ISM
and on the progenitor mass, the percentage of surviving dust ranges from 
$16\%$ to $100\%$. Also, compared to the fully mixed models, we note that 
in each case dust is destroyed to a minor degree.
In fact, in all the three unmixed models, most of the dust mass is produced
in the innermost layer (layer A). At the time the reverse shock reaches this region
of the ejecta, density is low and the effect of the reverse shock is less
destructive. In Table \ref{rev_unmixed} we note that, while layer B and C are highly 
affected by the passage of the reverse shock, dust in layer A survives in large quantities
and represents almost the totality of the survived dust mass. These calculations have 
been performed with an updated version of the code by Bianchi \& Schneider (2007) which 
considers also stratified ejecta, taking into account the gas (density, temperature, elemental abundances) 
and dust (composition, size distribution) properties in each shell depending on 
the position within the ejecta.

These results suggest that the effect of the reverse shock 
must be taken into account in order to have a reliable estimate of the final 
dust mass and composition.

\section{PopIII faint SN progenitors}
We know from observations that a fraction of $\sim 10\%-20\%$ of iron-poor 
stars observed in the Galactic Halo with $\rm [Fe/H]<-2$ are carbon-rich 
(Yong et al.~2013; Placco et al. 2014 and references therein).  Most notably,
the observed frequency of CEMP stars that do not show overabundances of 
neutron-capture elements, the so-called CEMP-no stars, increases to $80\%$
at $\rm [Fe/H] < -4$, with interesting implications for chemical evolution models
and the formation pathway of these hyper-iron poor stars (de Bennassuti et al. 2014).

Different scenarios have been proposed to explain the observed elemental abundances
of CEMP-no stars (see Norris et al. 2013 and references therein). Among these, 
stellar winds of fast-rotating massive stars (Maeder, Meynet \& Chiappini 2015),
a single Pop~III supernova, experiencing mixing and fallback 
after the explosion (Umeda \& Nomoto~2002; Iwamoto et al.~2005; Tominaga et al.~2007; 
Tominaga~2009; Heger \& Woosley~2010; Tominaga et al.~2014; Yong et al.~2013; Ishigaki 
et al.~2014; Marassi et al.~2014), or in terms of an almost failed Pop~III SN (with a 
large fallback) exploding in an environment pre-enriched by one or more normal Pop~III 
supernovae (Limongi et al.~2003). 

In Marassi et al.~(2014) we intepreted the surface 
element abundances of SMSS J031300, the recently discovered CEMP-no stars with 
[Fe/H]$<-7.1$ (Keller et al.~2014), in the framework of the mixing-fallback scenario. 
In particular, in that paper dust formation in the ejecta of Pop III faint 
SNe was investigated for the first time, showing that - depending on the extent of mixing experienced 
by the ejecta and on the partial destruction by the SN reverse shock - between
$0.025$ and $\rm 2.25 \,M_{\odot}$ of carbon dust forms. These dust masses are large 
enough to activate dust-driven fragmentation (Schneider et al.~2012) in the parent 
star-forming cloud of SMSS J031300, even accounting for the dilution and mixing of the
SN ejecta with the surrounding pristine gas (Marassi et al.~2014).

In this section, we use the same procedure applied in Marassi et al.~(2014) 
to estimate the mass and composition of dust that forms in the ejecta of 
faint Pop~III SNe. The SN models have been selected by
comparing the predicted elemental yields with the observed surface elemental
abundances of the four C-enhanced, hyper iron-poor stars currently known, namely:
HE1327-2326 (Frebel et al.~2005), HE0107-5240 (Christlieb et al.~2002), 
HE0557-4840 (Norris et al.~2007) and SMSS J031300 (Keller et al.~2014). For 
each of these, we vary the mixing and fallback efficiencies so as to minimize the 
scatter between the observed abundance pattern and the elemental yields predicted
by Pop~III SN models (Limongi \& Chieffi 2012)\footnote{In Marassi et al. (2014)
we estimate that in less than 1 Myr the ejecta material can be well mixed and diluted with 
pristine gas, enriching the gas cloud out of which second generation C-enhanced stars form
with a [Fe/H] consistent with the observed values.}

\begin{figure*}
\includegraphics[width=8.50cm]{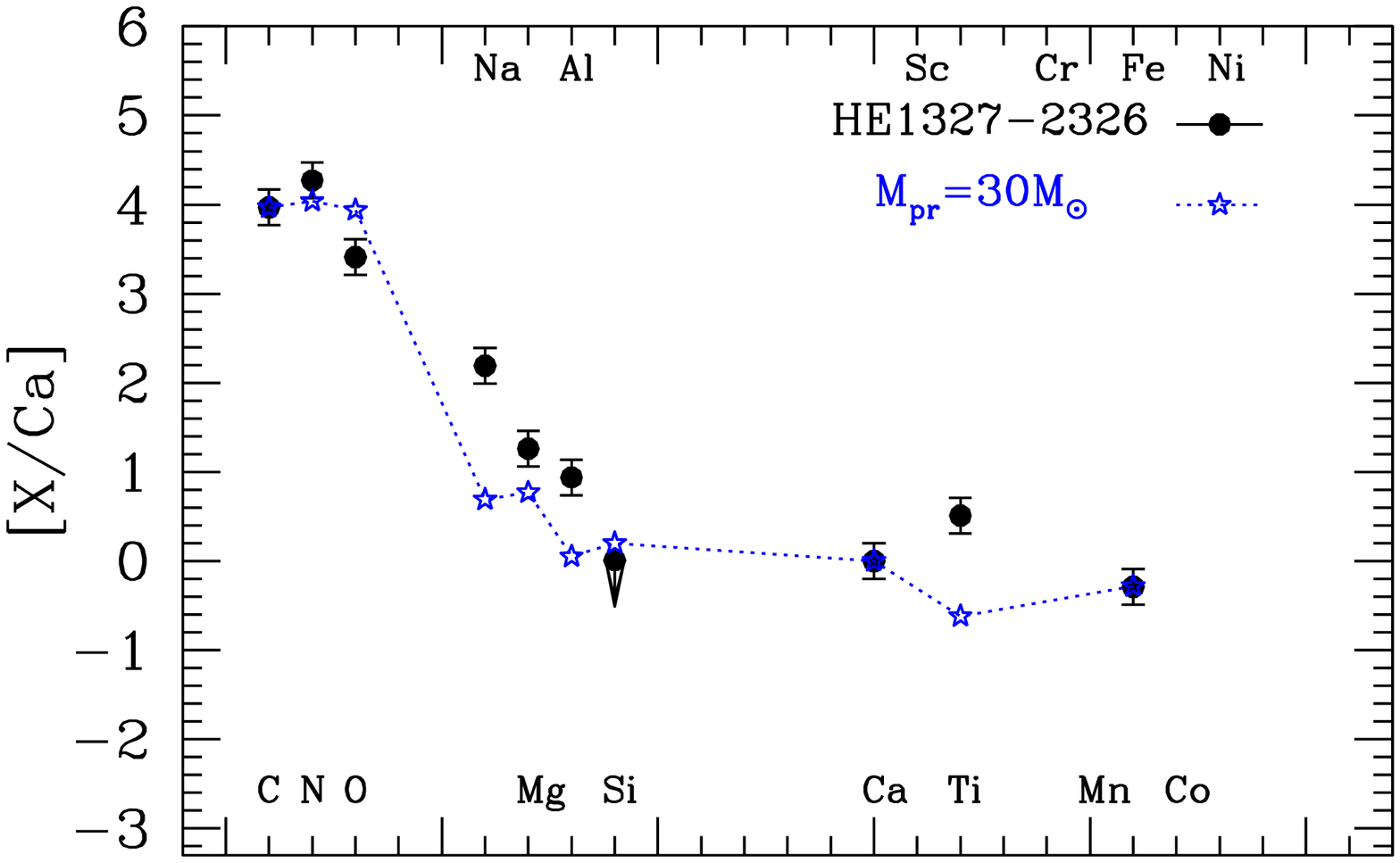}
\hspace{-1.7cm}
\vspace{-3.75cm}
\includegraphics[width=8.50cm]{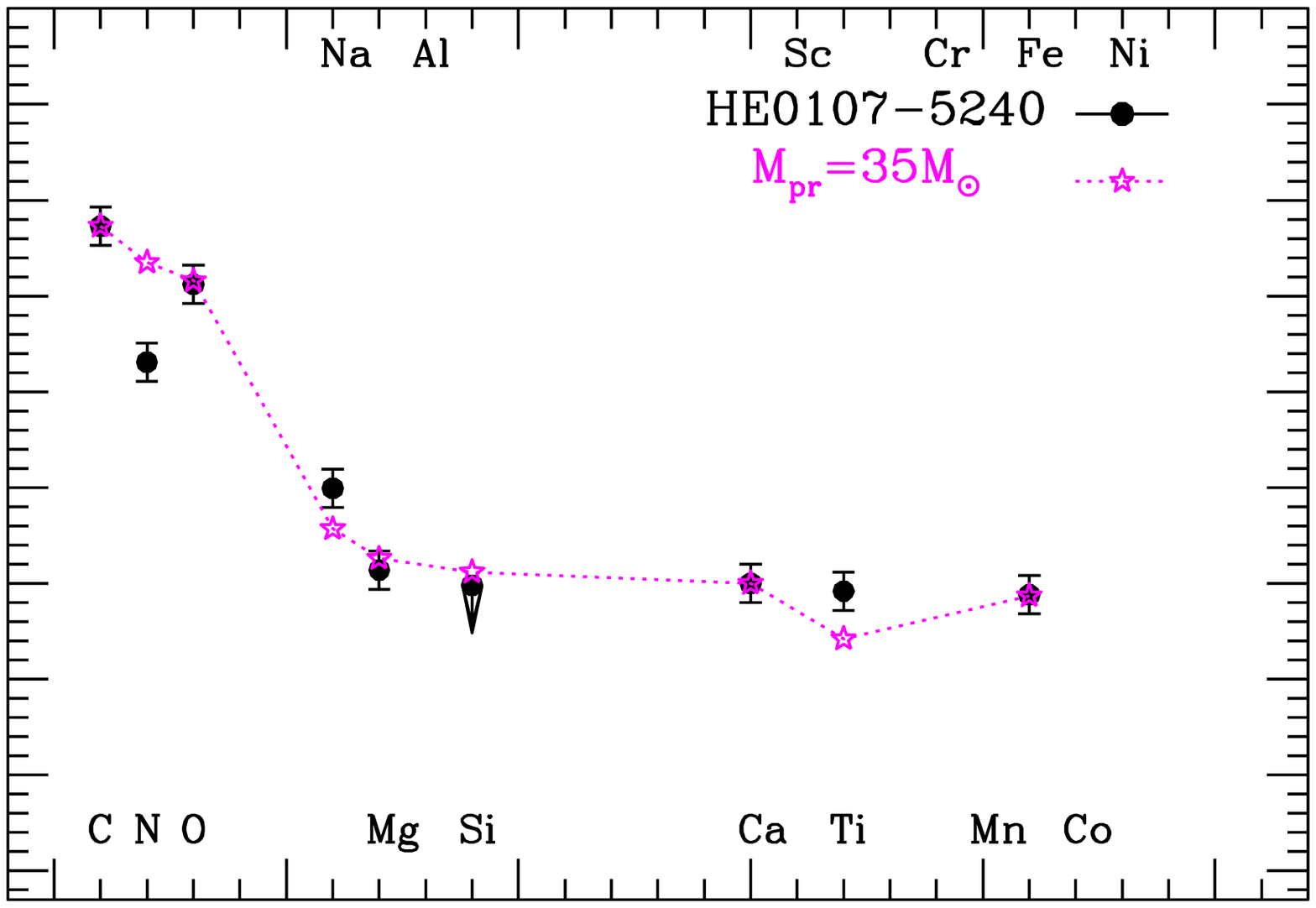}
\includegraphics[width=8.50cm]{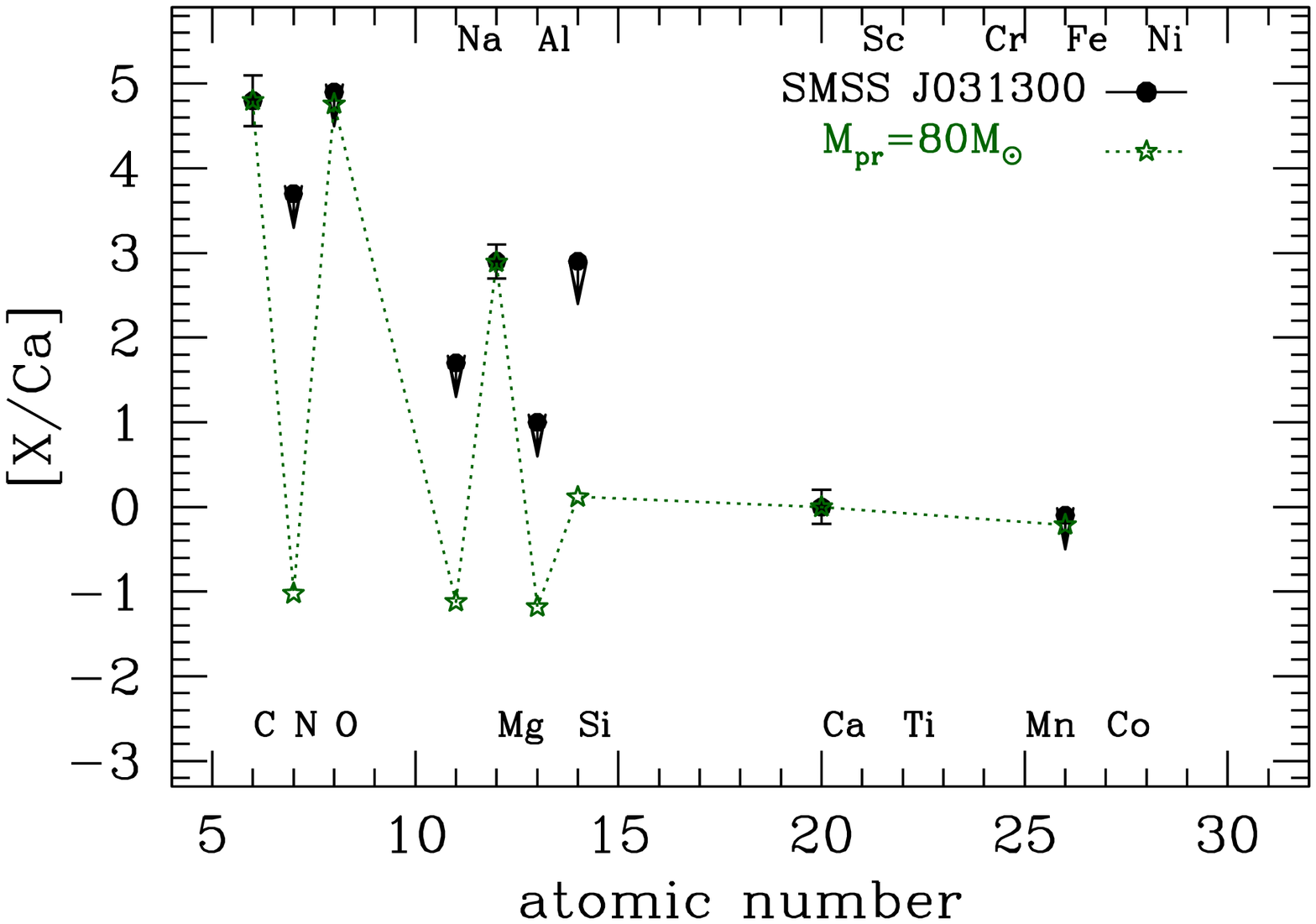}
\hspace{-1.7cm}
\includegraphics[width=8.50cm]{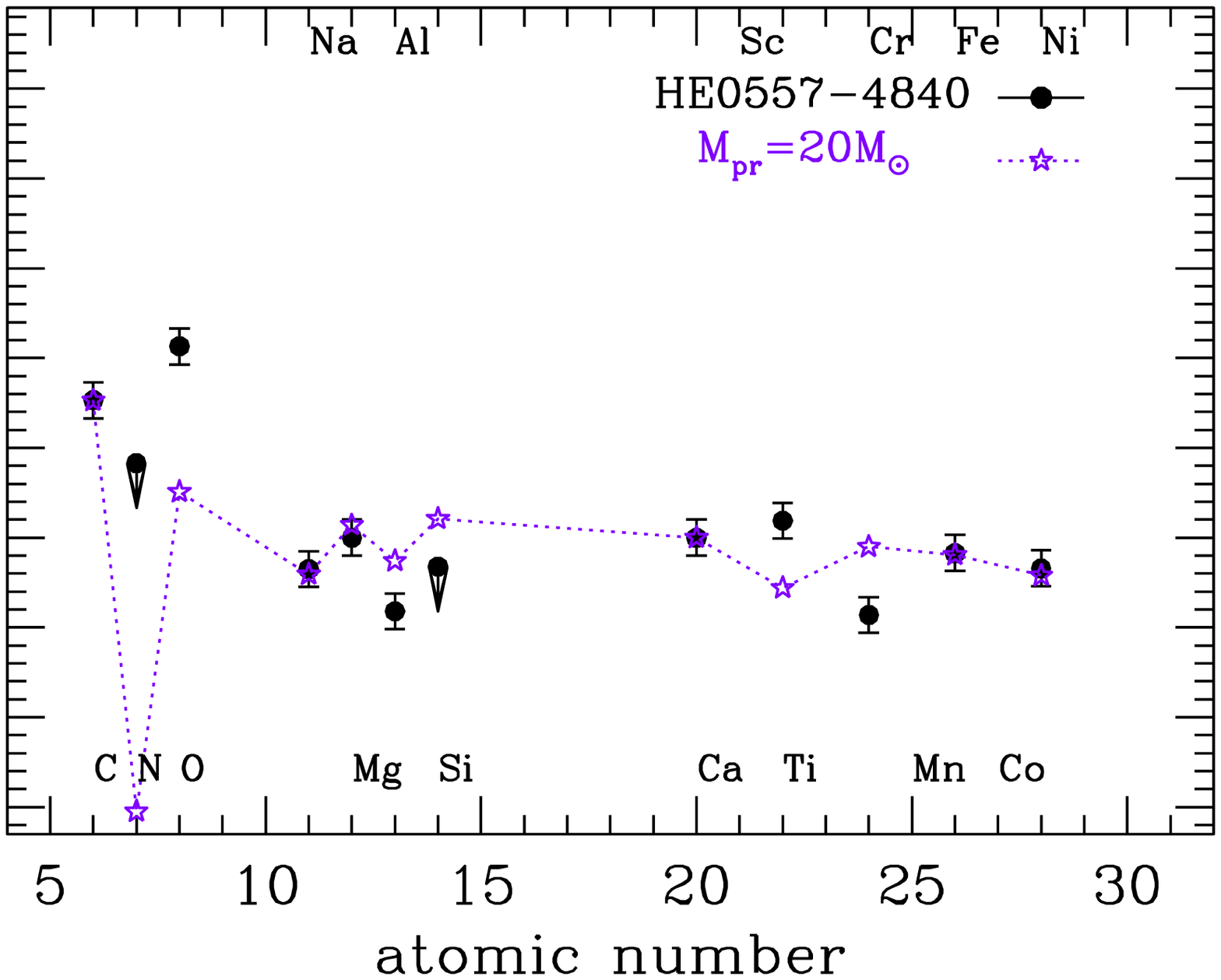}
\caption{Comparison between the observed elemental abundance 
ratios of the CEMP-no stars and the chemical yields of Pop~III 
faint SN with progenitor masses of $\rm 30, 35, 20, and \,80~M_{\odot}$. 
Mixing and fallback are chosen so as to minimize the scatter 
with the observations (black points). Dots with arrows show 
upper limits and filled points with
errorbars indicate the detections.}
\label{HYP}
\end{figure*}

The surface chemical abundances of HE1327-2326, HE0107-5240, and HE0557-4840  
have been taken from the recent compilation by Norris et al.~(2013)  (see their Table 4; here
we normalize to the solar abundances of Asplund et al.~2009), which are based on high-resolution data, but 
are determined using 1D, LTE model-atmosphere analysis. The upper limit on silicon 
for HE1327-2326, HE0107-5240 and HE0557-4840 is taken from Yong et al.~(2013).
For consistency,  we consider the observed elemental abundances of SMSS J031300 as
derived from 1D model atmosphere (Marassi et al.~2014).  
In Fig.~\ref{HYP} we show the comparison between the observed elemental 
abundances and the best fit models for each of the four stars. We find
that the data are reproduced by Pop~III SNe with progenitor masses in 
the range $\rm [20 - 80] \,M_{\odot}$ that experience strong fallback.  
For all the models we have 
identified the mass-coordinates that defines the extent of mixing and 
fallback that better reproduce the observed abundaces, with particular 
attention on [C/Ca], [Mg/Ca] and [O/Ca] - that are extremely important 
for dust formation - without exceeding the upper limits on [Fe/Ca].  
In the following, we discuss the properties of the faint Pop~III SN progenitors inferred
from the observed abundances of each star, which we also report in Table 9.

For {\bf HE1327-2326}, nitrogen has been detected and the observed 
abundance shows comparable value of [N/Ca] and [C/Ca]. Since a 
substantial amount of N is produced only by Pop III SN models with progenitor 
masses in the range $\rm [25 - 35]\,M_\odot$ (see Limongi \& Chieffi~2012 
for more details), we have limited the exploration of the mixing and
fallback procedure to this mass range. The best agreement for HE1327-2326 
is provided by the $\rm 30 \, M_{\odot}$ progenitor model with a mass-cut of $\rm 5\, M_\odot$
and an ejected mass of $\rm M(^{56}Ni)$ of $\rm 3.68\times 10^{-7} \, M_\odot$. Note that a [Si/Ca] ratio close to 
the observed upper limit implies a substantial under-production of 
[(Na,Mg,Al)/Ca] ratios. Since the amount of Si in the ejecta plays 
a crucial role in the dust formation process, we chose to fit the [Si/Ca] 
ratio, even if the abundances of Na, Mg and Al are underproduced. 

The same procedure has been applied to {\bf HE0107-5240} finding the 
best agreement for a SN progenitor model with $\rm 35 \, M_{\odot}$ a mass-cut of $\rm 8 M_{\odot}$
and a $\rm M(^{56}Ni)$ mass of $\rm 9.12\times 10^{-7} \, M_\odot$, which reproduces 
the observed [C/Ca] and [O/Ca] and predicts a [Na/Ca] ratio close to the 
observed one. 

For {\bf HE0557-4840} we have searched for progenitors which favors low 
[N/Ca] due to the low upper-limit inferred from observations. We 
select a $\rm 20 \, M_{\odot}$ progenitor with $\rm M_{cut} = 5 \, M_\odot$ and $\rm M(^{56}Ni) = 9.19\times10^{-7} \,M_\odot$, 
that reproduces also
the observed [C/Ca] and [Mg/Ca] abundances. We note that none of the models is able to 
reproduce, at the same time, the observed [O/Ca] and [Mg/Ca] abundances 
(see also Limongi \& Chieffi~2012 and Ishigaki et al.~2014), particularly 
the low-mass progenitors, which have lower value of [Mg/O] in the ejecta. 
\begin{figure}
\includegraphics[width=8.50cm]{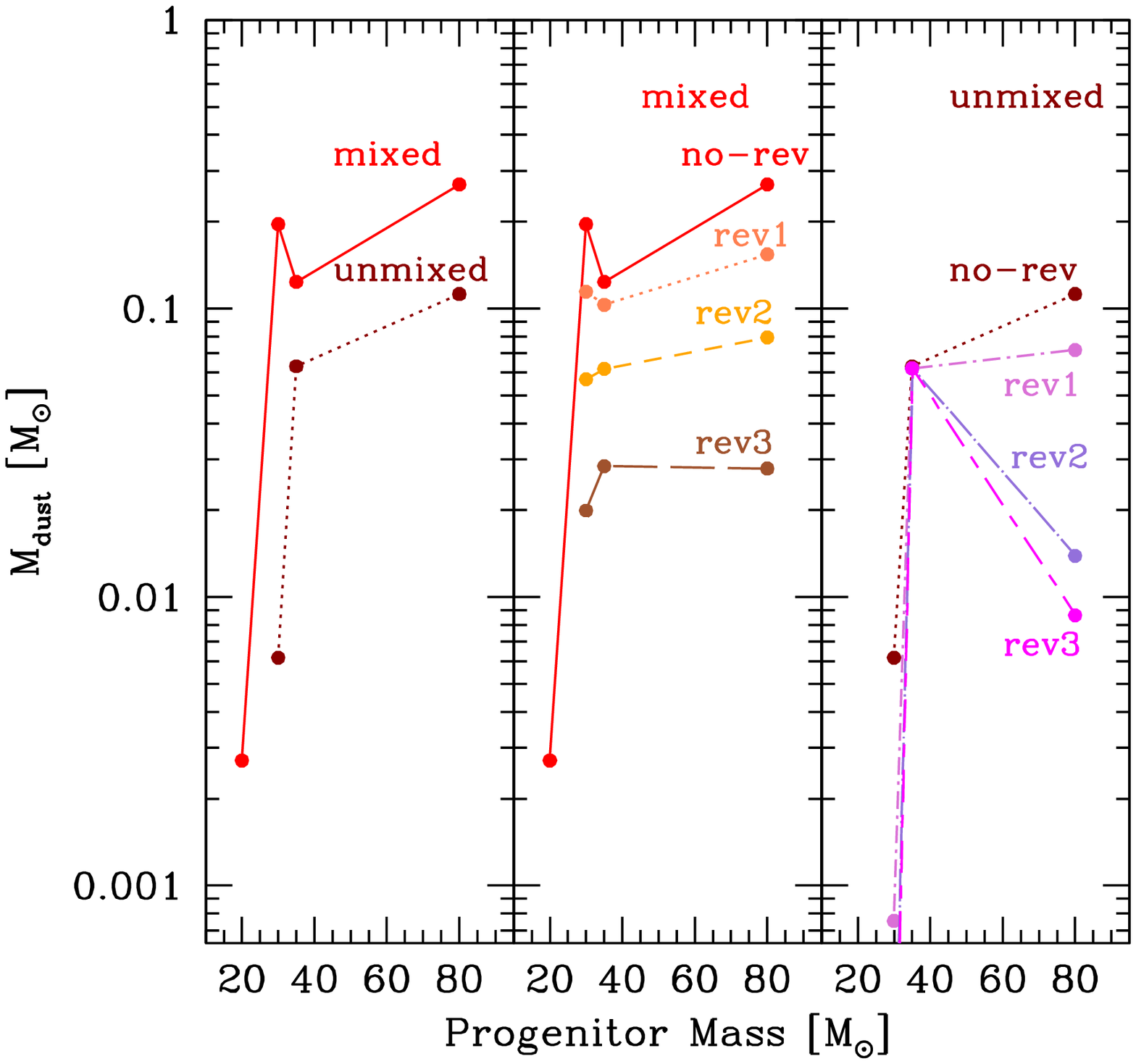} 
\caption{Left panel: the mass of dust at the end of nucleation for mixed and unmixed ejecta for the four faint SN progenitors. Central panel: 
the mass of dust at the end of nucleation and after the passage of a reverse shock of increasing intensity for the four faint SN progenitors 
with mixed ejecta considered in this study. Right panel: the same as in the central panel but for faint SN models with
stratified ejecta. 
From top to bottom: no reverse shock models 
({\bf norev}), models with $\rm n_{ISM} = 0.06 \, ({\bf rev1}), 0.6 \,({\bf rev2}) 
and\, 6 \, \rm cm^{-3}$ ({\bf rev3})}. 
\label{HYP_rev_shock}
\end{figure}

Finally, for {\bf SMSS J031300} we have reported only the best-fit obtained for 
the $\rm 80 \,M_{\odot}$ model ($\rm M_{cut} = 24 \, M_{\odot}$ and $\rm M(^{56}Ni) = 1.43\times10^{-7} \, M_\odot$) 
selected by Marassi et al.~(2014), to which we refer 
the reader for a detailed discussion.
\section{Dust yields from faint Pop III SN}
This section presents the results of the dust formation calculation for 
Pop~III faint SN progenitors. To enable a comparison with the results 
by Marassi et al.~(2014), we first discuss 
the results for uniformly mixed ejecta, which are presented in 
Table~\ref{tabappIII:faint} in the Appendix, where we give - for each of 
the four selected faint SN progenitors - the explosion parameters, the
thermodynamical properties, the metal yields of the key elements 
in the nucleation process, the total amount of dust and the masses 
of molecules. 
Similarly to Marassi et al.~(2014) we find that for all the progenitor models 
investigated, the only grain species that is able to condense and grow is 
amorphous carbon. This is a consequence of the extensive fallback, 
which determine an ejecta composition dominated by carbon atoms, with 
an amount of Mg, Si and Al which is too low to enable 
the condensation of other grain species (Marassi et al.~2014). 
We find that the molecular reactions involving SiO are negligible.  
The mass of AC is in the range $\rm [2.7\times 10^{-3} - 0.269]\,M_{\odot}$, 
depending on the initial carbon abundance and on the mean ejecta density 
and temperature evolution. Due to the interplay of these quantities it is 
not possible to establish a direct correspondence between progenitor 
mass, ejected mass and the dust mass produced. As shown in Marassi 
et al.~(2014), AC grains do not forms if the formation channels of carbon 
monoxide are efficient in subtracting carbon atoms from the ejecta, 
and this efficiency is greater when the mean ejecta density is larger. 
As a consequence, we note that HE0557-4840 have the greater condensation 
efficiency but the smaller AC mass produced. Equally important is the 
temperature evolution which, in the case of HE0107-5240, enables the 
formation of $\rm C_{2}$, subtracting C atoms from the ejecta. 

In faint SNe, mixing occurs due to Rayleigh-Taylor instabilities up to 
a mass-coordinate that is very close to the mass-cut (Umeda \& Nomoto 2002); 
as a consequence, the material beyond the mass-cut is likely to be stratified, 
and dust nucleation in unmixed ejecta gives a more reliable estimates of the 
total mass of dust produced. Table~\ref{tabappIII:faint_unmixed} shows the 
results for the unmixed ejecta model. For three of the four progenitor models, 
the stratified ejecta is composed of two layers A and B that have different 
average temperature, density and chemical composition. In 
particular, in all three unmixed models analyzed, the internal layers A 
have mean ejecta density larger compared to the external layers B, enabling 
a very efficient formation of CO which inihibits the formation of AC.

The consequence of stratification is a reduction of the total mass of dust 
produced, that ranges between $\rm [7.5\times 10^{-4} - 0.11]\,M_{\odot}$.
In the left panel of Fig.~{\ref{HYP_rev_shock}},  we plot 
the total mass of dust before the passage of the reverse shock for mixed and unmixed models.
Similarly to what done in Section 4, we have studied the impact of the 
reverse-shock on the above dust masses. 
The grain sizes originally follow a log-normal 
distribution in the range $\rm [0.03 - 0.22]\mu m$. As expected, 
after the passage of the shock the distribution flattens and shifts to lower grain sizes, 
due to the erosion of the larger grains caused by sputtering. 
 In the central and right panels of Fig.~{\ref{HYP_rev_shock}}, 
we show the mass of dust that survives in uniformly mixed and stratified ejecta
models, respectively. We find that, depending on the progenitor model, 
dust can be totally destroyed (as in the case of $\rm HE0557-4840$) or that
between 10  to 80$\%$ of the original mass can survive. 
The passage of reverse shock in the unmixed case is less destructive than in the case of
the fully mixed model. As discussed in Section 4, dust in external layers is highly affected
by the reverse shock. On the contrary dust produced in internal layers, as in the case 
of $\rm HE0107-5240$, is able to survive in large quantities, finding itself in milder conditions 
compared to the external layers.

\begin{figure*}
\includegraphics[width=8.50cm]{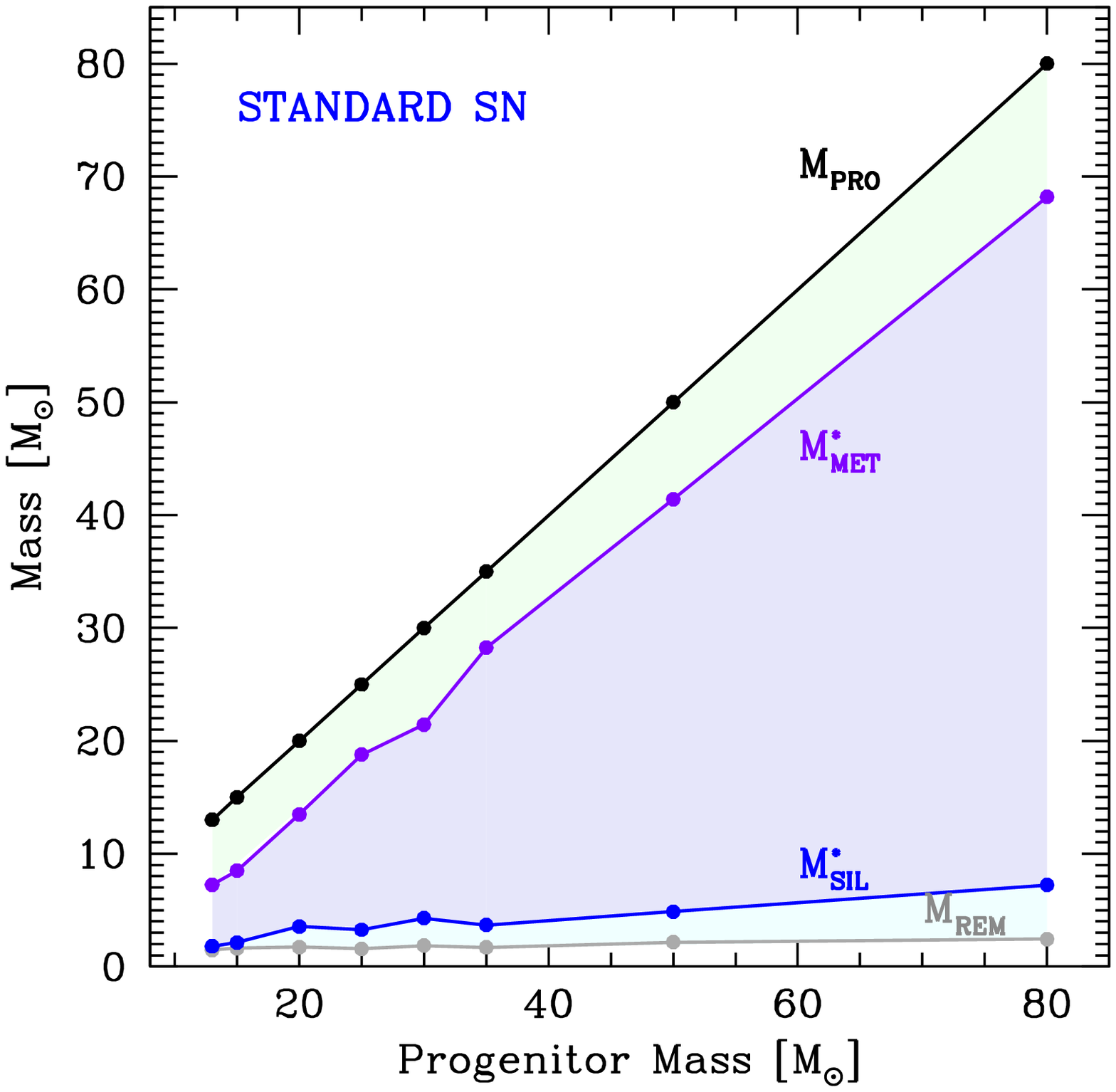}
\includegraphics[width=8.50cm]{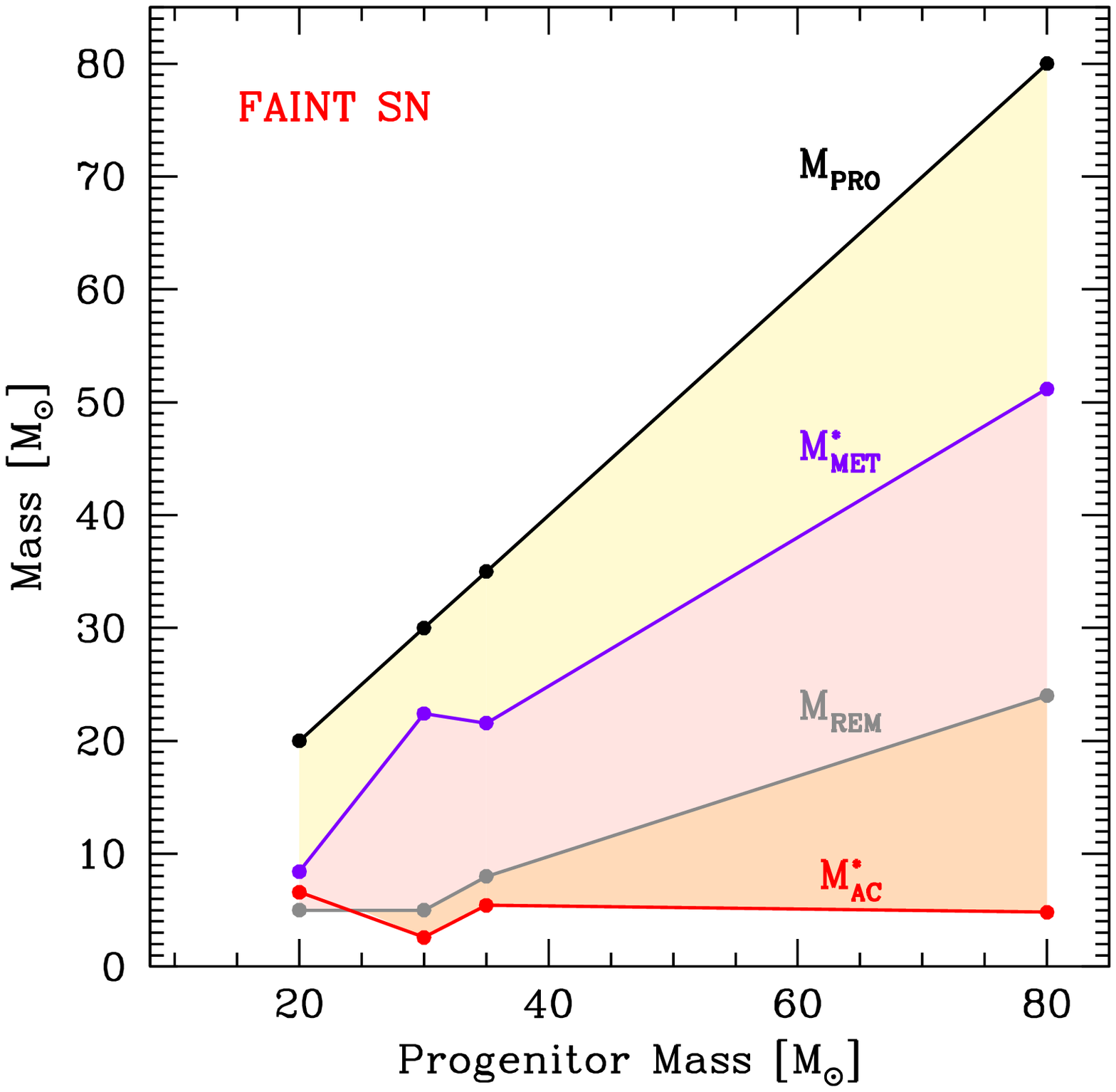}
\caption{Mass of the SN progenitor, of the remnant, of metals and dust mass as a function of the progenitor mass for 
 {\it standard}  (left panel) and {\it faint} (right panel) Pop~III SNe. Note that the mass of metals and
 dust have been multiplied by $\rm M_{eje}/(M_{met}+M_{dust})$ (see text).}
\label{cc_faint}
\end{figure*}
\section{Conclusions}
In this study we have estimated the dust mass produced in the early 
Universe by Pop~III {\it standard} and {\it faint} core-collapse 
SNe. For {\it standard} core-collapse events, which do not experience 
a strong fallback, the metal-free progenitors are calibrated to reproduce 
the average elemental abundances of Galactic halo stars with
$\rm [Fe/H] < -2.5$. We have enlarged our study to include Pop~III faint SNe, which 
are fine-tuned to reproduce the abundance patterns observed on the surface 
of the sub-sample of carbon-enhanced hyper iron poor stars.

To estimate the dust mass produced that survives the subsequent 
impact of the reverse shock, we have updated the previously developed 
code of Bianchi \& Schneider (2007) to include formation/destruction 
processes of CO, SiO, $\rm C_{2}$ and $\rm O_{2}$ molecules that are 
crucial in subtracting gas-phase elements from the ejecta. We have 
investigated how our poor knowledge of mixing can affect the 
resulting dust yields by analyzing both uniformly mixed and 
stratified ejecta. Finally, to have a more realistic estimate 
of the dust mass that will be injected into the ISM, we have 
analized the effect of the passage of the reverse-shock. 
Our main results are that:
\begin{itemize}
\item Standard core-collapse Pop~III SNe are efficient dust producers when the 
ejecta are uniformly mixed; the newly formed dust mass ranges between 
[0.18 - 3]\, M$_{\odot}$ and is dominated by silicates. 
Depending on the density of the interstellar medium, the impact 
of the reverse shock on the newly formed dust is to reduce the original
dust mass by a factor which ranges between $3\%$ to $48\%$, changing the
relative abundance of different dust species. The mass 
of dust produced depends on the degree of mixing experienced by the 
ejecta. In general, silicates, magnetities and alumina grains are 
formed in the internal layers, while amorphous carbon forms only in 
the external layers that are carbon and oxygen dominated. In 
stratified ejecta dust does not form at the same time in each layer, with
silicates forming in the inner layers before AC in the external ones.
For the same SN progenitor model, the mass of dust formed in unmixed ejecta
can be smaller or higher than in the corresponding mixed case, with variations
which amount to $3 - 10\%$. 

\item We confirm the results found by Marassi et al.~(2014) that dust can form
in the ejecta of faint Pop~III SNe. As a consequence of 
the larger fallback experienced with respect to standard core-collapse Pop~III SNe, 
the ejecta is lacking in silicate, magnesium and alumina and the only dust grain that forms 
is amorphous carbon. The total amount of carbon dust produced in the 
uniformly mixed case varies between [$0.2-0.27$]M$_{\odot}$, and in the 
stratified case between [$7.4\times 10^{-4}-0.11$]M$_{\odot}$ depending 
on the model. After the passage of the reverse shock, the dust that 
survives ranges between $10\%$ to $80\%$.

The mass and composition of dust formed in the ejecta 
of Pop~III SNe depends on the degree of fallback. In {\it standard} Pop~III 
SNe, the mass-cut corresponds to a mass coordinate in the range $\rm [1.48 - 2.44] \,M_{\odot}$. 
As a result, almost all the mantle is ejected with a greater amount of internal 
elements, such as silicon and magnesium, favoring the condensation of silicates. 
The observed elemental abundances of CEMP stars with $\rm [Fe/H] < -4$ 
require Pop~III SN progenitors with masses in the range $[20 - 80]\, M_{\odot}$
which experience a strong fallback, with mass-cut which 
correspond to mass coordinates in the range $\rm [5 - 24]\,M_{\odot}$. 
These {\it faint} SNe leave a much larger remnant behind and eject a smaller amount
of metals, with mantles dominated by light elements. 
These properties of the SNe largely affect the mass and composition of dust which
condense in the ejecta, as shown by Fig.~\ref{cc_faint}, where we 
summarize our results showing the mass of metals and of the dominant grain species 
produced by {\it standard} (left panel) and {\it faint} (right panel) Pop~III SNe.
Note that for each SN model, we have rescaled the mass of metals and dust as
$\rm M_{met}^* = M_{eje} M_{met}/(M_{met}+M_{dust})$ and 
$\rm M_{dust}^* = M_{eje} M_{dust}/(M_{met}+M_{dust})$, to clearly display these physical quantities.

Finally we have constructed, starting from our set of metal-free 
progenitors, a dust data grid for Pop III SNe which provides the initial
conditions required to simulate the properties of the first metal-enriched 
star forming regions at high-redshift. In addition, we plan to extend this
analysis to higher metallicity to incorporate these new mass- and metallicity-dependent 
yields in chemical evolution models with dust.

\end{itemize}
\section*{Acknowledgments}
We thank the anonymous Referee for her/his careful reading of the manuscript and 
useful comments. 
The research leading to these results has received funding from the European Research Council 
under the European Union’s Seventh Framework Programme (FP/2007-2013)/ERC Grant Agreement 
n. 306476. We aknowledge financial support from PRIN MIUR 2010-2011, project 
``The Chemical and dynamical Evolution of the Milky Way and Local Group Galaxies'', 
prot. 2010LY5N2T

\section*{Appendix: Final dust grids and tables}
In this appendix we report all the useful tables: for all progenitors 
we report chemical composition (yields), mass of molecules, mass of dust etc.
\begin{table*}
\caption{Properties of the Pop~III SNe, including the explosion energy [$10^{51}$~erg], the ejecta velocity [$\rm km~s^{-1}$], the mass of the ejecta, 
the mass cut and the mass of the helium core [$M_\odot$], the gas number density [$\rm cm^{-3}$] and the radius of He core, $\rm R_0 $ [cm] 
at $\rm t=t_{ini}$ [sec] when adiabatic expansion starts (see text); the initial masses of C, O, Mg, Si, Al, Fe, and $^{56}$Ni which decays in $^{56}$Co fueling $\gamma$-luminosity (see eq.~\ref{Lgamma}), the mass of molecules, CO, SiO, $\rm O_{2}$, $\rm C_{2}$ and the grains formed in the 
expanding ejecta [$\rm M_{\odot}$]. Each model name identifies the progenitor mass.}
\label{tabappI:masses_normal}
\begin{tabular}{@{}lcccccccc}
\hline
\multicolumn{9}{c}{Pop~III SN - Fully Mixed Ejecta Models} \\
\hline
       & $\rm 13 M_{\odot}$ &  $\rm 15 M_\odot$ & $\rm 20 M_\odot$ & $\rm 25 M_\odot$  & $\rm 30 M_\odot$  &$\rm 35 M_\odot$ &$\rm 50 M_\odot$&$\rm 80 M_\odot$ \\
\hline
 $\rm E_{exp}$ &$0.5$& $0.7$ & $1.0$ & $1.1$ & $1.6$ &$1.3$ & $2.6$ & $5.2$\\
 $\rm v_{eje}$&$2696$&$2960$&$3027$ &$2805$&$3086$&$2557$&$3017$&$3351$\\ 
 $\rm M_{eje}$&$11.52$&$13.38$&$18.28$&$23.42$&$28.14$&$33.30$&$47.84$&$77.55$\\
 $\rm M_{cut}$&$1.48$&$1.62$&$1.72$&$1.58$&$1.86$&$1.70$&$2.16$&$2.44$\\
 $\rm M_{He_{core}}$&$2.81$&$3.40$&$5.18$&$5.37$&$7.22$&$11.13$&$18.11$&$31.94$\\ 
\hline
$\rm R_{0}$& $2.65\times 10^{14}$&$2.44\times 10^{14}$&$1.47\times 10^{14}$&$3.75\times 10^{14}$&$4.68\times 10^{14}$& $1.98\times 10^{14}$&$2.51\times 10^{14}$&$4.78\times 10^{14}$\\
$\rm t_{ini}$& $2.11\times 10^{6}$&$1.78\times 10^{6}$&$1.58\times 10^{6}$&$3.72\times 10^{6}$&$3.96\times 10^{6}$&$3.11\times 10^{6}$&$3.12\times 10^{6}$&$4.14\times 10^{6}$\\
$\rm n_{0}$& $2.71\times 10^{11}$&$7.18\times 10^{11}$&$8.34\times 10^{12}$& $9.31\times 10^{11}$&$7.39\times 10^{11}$&$1.26\times 10^{13}$&$1.24\times 10^{13}$&$3.65\times 10^{12}$\\
\hline
 $\rm M(^{56}Ni)$&$2.20\times 10^{-2}$&$3.85\times 10^{-2}$&$7.12\times 10^{-2}$&$9.54\times 10^{-2}$&$2.10\times 10^{-1}$&$1.81\times 10^{-1}$&$2.86\times 10^{-1}$&$4.86\times 10^{-1}$\\ 
 $\rm M_{C}$ &  $7.52\times 10^{-2}$&$1.56\times 10^{-1}$&$2.99\times 10^{-1}$&$4.48\times 10^{-1}$&$7.02\times 10^{-1}$&$6.21\times 10^{-1}$ &$1.44$ & 2.43\\        
 $\rm M_{O}$ &  $1.38\times 10^{-1}$&$2.91\times 10^{-1}$&$9.31\times 10^{-1}$ & $1.96$ & $2.96$ &$4.35$ &$8.60$ & $18.22$\\       
 $\rm M_{Mg}$ & $1.83\times 10^{-2}$&$3.61\times 10^{-2}$&$9.99\times 10^{-2}$ & $9.34\times 10^{-2}$ & $1.33\times 10^{-2}$ &$2.15\times 10^{-1}$ &$2.03\times 10^{-1}$&$3.66\times 10^{-1}$\\      
 $\rm M_{Si}$ & $4.42\times 10^{-2}$&$8.49\times 10^{-2}$&$1.12\times 10^{-1}$ & $1.59\times 10^{-1}$&$ 3.23\times 10^{-1}$ &$1.72\times 10^{-1}$ &$4.46\times 10^{-1}$&$8.22\times 10^{-1}$\\    
 $\rm M_{Al}$ & $6.63\times 10^{-4}$&$1.42\times 10^{-3}$&$3.11\times 10^{-3}$ &$2.50\times 10^{-3}$&$ 2.31\times 10^{-3}$&$3.49\times 10^{-3}$&$1.85\times 10^{-3}$ &$3.61\times 10^{-3}$\\        
 $\rm M_{Fe}$ & $2.31\times 10^{-2}$& $4.03\times 10^{-2}$&$7.39\times 10^{-2}$ & $1.00\times 10^{-1}$& $2.18\times 10^{-1}$ &$1.87\times 10^{-1}$&$2.95\times 10^{-1}$&$5.04\times 10^{-1}$\\       
\hline
 $\rm M_{CO}$ &$9.60\times 10^{-3}$ & $3.43\times 10^{-2}$& 0.65 &0.896 &1.25 &1.45 & 3.36 &5.66\\ 
 $\rm M_{SiO}$&-&-&-&-&-&-&-&-\\
 $\rm M_{O_2}$&-&-&-&-&-&$0.77$&$1.41$&$1.64$\\
 $\rm M_{C_2}$&-&-&-&-&-&-&-&-\\
\hline
 $\rm M_{AC}$&$7.10\times 10^{-2}$&$0.14$& $1.80\times 10^{-2}$&$6.36\times 10^{-2}$&$0.16$&-&- &-  \\
 $\rm M_{MgSiO_3}$&$2.17\times 10^{-4}$&$5.10\times 10^{-5}$&-&$3.86\times 10^{-4}$&$5.62\times 10^{-3}$&-&$3.23\times 10^{-6}$ & $5.76\times 10^{-4}$\\
 $\rm M_{Mg_2SiO_4}$&$5.28\times 10^{-2}$&$0.104$&$0.29$&$0.27$&$0.38$&$0.62$&$0.59$&$1.06$\\
 $\rm M_{Fe_3O_4}$&$3.00\times 10^{-2}$& $5.65\times 10^{-2}$&$0.11$&$0.14$&$0.30$&$0.263$&$0.41$&$0.70$\\
 $\rm M_{SiO_2}$&$2.12\times 10^{-2}$& $4.93\times 10^{-2}$&$0.11$&$0.21$&$0.46$& $0.102$&$0.70$ & $1.31$ \\
 $\rm M_{Al_2O_3}$&$1.25\times 10^{-3}$&$2.67\times 10^{-3}$ &$5.88\times 10^{-3}$ &$4.72\times 10^{-3}$&$4.36\times 10^{-3}$&$6.60\times 10^{-3}$&$3.50\times 10^{-3}$& $6.83\times 10^{-3}$\\
\hline
 $\rm M_{dust}$& $0.176$ & $0.35$ &  $0.54$ & $0.68$ &  $1.32$ &  $0.99$ &  $1.71$ & $3.07$\\
\hline
\end{tabular}
\end{table*}
\begin{table*}
\caption{Properties of the {\rm 15\Msun} Pop~III SN unmixed ejecta model. In unmixed models $\rm R_0$ and 
$\rm n_0$ are the mean radius and density of the layers.}
\label{tabappII:masses_unmixed_15}
\begin{tabular}{@{}lccc}
\hline
\multicolumn{4}{c}{$15M_\odot$ Unmixed Ejecta Model} \\
\hline
 $\rm E_{exp}$ =0.7 & $\rm M_{He_{core}}$ = 3.40 & $\rm v_{eje}$ = 2960 & $\rm M_{eje}$ = 13.38 \\
\hline
 & zone~A $(1.62-1.98)$\Msun & zone~B $(1.98-2.22)$\Msun & zone~C $(2.22-3.40)$\Msun \\
\hline
$\rm R_{0}$& $3.50\times 10^{14}$ & $1.96\times 10^{14}$ &$2.22\times 10^{14}$\\
$\rm t_{ini}$& $3.19\times 10^{5}$ & $1.77\times 10^{6}$ &$1.78\times 10^{6}$\\
$\rm n_{0}$& $7.18\times 10^{12}$ & $7.50\times 10^{12}$ &$2.37\times 10^{11}$ \\
\hline
$\rm M(^{56}Ni)$&$3.85\times 10^{-2}$&-&-\\
$\rm M_{C}$&$1.88\times 10^{-3}$ & $9.48\times 10^{-2}$ & $5.95\times 10^{-2}$\\
$\rm M_{O}$&$0.17$&$0.12$&$7.50\times 10^{-3}$\\
$\rm M_{Mg}$&$3.52\times 10^{-2}$&$8.87\times 10^{-4}$&-\\
$\rm M_{Si}$&$8.45\times 10^{-2}$&$3.60\times 10^{-4}$&-\\
$\rm M_{Al}$&$1.38\times 10^{-3}$&$3.36\times 10^{-5}$&-\\
$\rm M_{Fe}$&$4.04\times 10^{-2}$&-&-\\
\hline
$\rm M_{CO}$&$4.38\times 10^{-3}$ & $0.18$ &$1.32\times 10^{-2}$\\
$\rm M_{SiO}$&- &- &-\\
$\rm M_{O_2}$&$3.00\times 10^{-2}$ &- &-\\
$\rm M_{C_2}$&- &- &$1.34\times 10^{-2}$\\
\hline
$\rm M_{AC}$&- & $1.69\times 10^{-2}$ &$4.03\times 10^{-2}$\\
$\rm M_{MgSiO_3}$&- &- &-\\
$\rm M_{Mg_2SiO_4}$&$0.10$ & $1.77\times 10^{-3}$ &-\\
$\rm M_{Fe_3O_4}$&$7.46\times 10^{-2}$ &- &- \\
$\rm M_{SiO_2}$&$0.14$ &- &-\\
$\rm M_{Al_2O_3}$& $2.61\times 10^{-3}$ & $6.36\times 10^{-5}$  &-\\
$\rm M_{Fe}$& $1.25\times 10^{-2}$&-&-  \\
\hline
$\rm M_{dust}$&0.33 & $1.88\times 10^{-2}$ &$4.03\times 10^{-2}$\\
\hline
\end{tabular}
\end{table*}
\begin{table*}
\caption{Properties of the {\rm 30\Msun} Pop~III SN unmixed ejecta model.}
\label{tabappII:masses_unmixed_30}
\begin{tabular}{@{}lccc}
\hline
\multicolumn{4}{c}{$30M_\odot$ Unmixed Ejecta Model} \\
\hline
 $\rm E_{exp}$ =1.6 & $\rm M_{He_{core}}$ = 7.22 & $\rm v_{eje}$ = 3086 & $\rm M_{eje}$ = 28.14 \\
\hline
 & zone~A $(1.86-3.99)$\Msun & zone~B $(3.99-5.20)$\Msun & zone~C $(5.20-7.22)$\\
\hline
$\rm R_{0}$&$3.29\times 10^{14}$&$3.63\times 10^{14}$& $4.52\times 10^{14}$   \\
$\rm t_{ini}$& $3.13\times 10^{6}$&$3.28\times 10^{6}$&$3.96\times 10^{6}$ \\
$\rm n_{0}$& $1.97\times 10^{12}$& $1.37\times 10^{13}$& $1.50\times 10^{12}$\\
\hline
$\rm M(^{56}Ni)$&$2.10\times 10^{-1}$&-&-\\ 
$\rm M_{C}$&$1.26\times 10^{-2}$&$0.23$&$0.45$\\
$\rm M_{O}$&$0.75$&$1.19$&$1.00$\\
$\rm M_{Mg}$&$9.91\times 10^{-2}$&$3.39\times 10^{-2}$&$1.94\times 10^{-5}$\\
$\rm M_{Si}$&$0.32$&$1.59\times 10^{-3}$&-\\
$\rm M_{Al}$&$1.65\times 10^{-3}$&$6.60\times 10^{-4}$&-\\
$\rm M_{Fe}$&0.22&-&-\\
\hline
$\rm M_{CO}$& $2.95\times 10^{-2}$& $0.53$& $1.04$ \\
$\rm M_{SiO}$&-&-&-\\
$\rm M_{O_2}$&$1.97\times 10^{-3}$& $0.46$&-\\
$\rm M_{C_2}$&-&-&-\\
\hline
$\rm M_{AC}$&-&-& $9.32\times 10^{-3}$\\
$\rm M_{MgSiO_3}$&-&-&-\\
$\rm M_{Mg_2SiO_4}$& $0.29$& $7.96\times 10^{-3}$&-\\
$\rm M_{Fe_3O_4}$&$0.30$&-&- \\
$\rm M_{SiO_2}$&$0.56$&-&-\\
$\rm M_{Al_2O_3}$& $3.11\times 10^{-3}$& $1.25\times 10^{-3}$&-\\
$\rm M_{Fe}$&- &-&-  \\
\hline
$\rm M_{dust}$&1.15&$9.21\times 10^{-3}$& $9.32\times 10^{-3}$\\
\hline
\end{tabular}
\end{table*}
\begin{table*}
\caption{Properties of the {\rm 50\Msun} Pop~III SN unmixed ejecta model.}
\label{tabappII:masses_unmixed_50}
\begin{tabular}{@{}lccc}
\hline
\multicolumn{4}{c}{$50M_\odot$ Unmixed Ejecta Model}\\
\hline
 $\rm E_{exp}$ =2.6 & $\rm M_{He_{core}}$ = 18.11 & $\rm v_{eje}$ = 3017 & $\rm M_{eje}$ = 47.84 \\
\hline
  &zone~A $(2.16-4.31)$\Msun & zone~B $(4.31-11.51)$\Msun& zone~C $(11.51-18.11)$\Msun\\
\hline
$\rm R_{0}$& $2.16\times 10^{13}$& $3.45\times 10^{13}$ & $1.11\times 10^{14}$\\
$\rm t_{ini}$&$4.76\times 10^{5}$  &$6.75\times 10^{5}$ &$3.12\times 10^{6}$\\
$\rm n_{0}$& $1.47\times 10^{14}$ & $1.24\times 10^{14}$ &$6.60\times 10^{12}$\\
\hline
$\rm M(^{56}Ni)$&$2.86\times 10^{-1}$&-&-\\ 
$\rm M_{C}$& $2.76\times 10^{-3}$ & $0.56$ & $0.87$  \\
$\rm M_{O}$& $1.05$ & $5.49$& $2.05$\\
$\rm M_{Mg}$& $9.56\times 10^{2}$ & $0.106$ & $6.60\times 10^{-4}$\\
$\rm M_{Si}$& $0.44$ & $5.40\times 10^{-3}$&-\\ 
$\rm M_{Al}$&-&$8.3\times 10^{-4}$&-\\
$\rm M_{Fe}$& $0.29$ &- &-\\
\hline
$\rm M_{CO}$& $6.44\times 10^{-3}$ & 1.31 &$1.76$\\
$\rm M_{SiO}$&- &- &-\\
$\rm M_{O_2}$&0.46 &3.46 &$1.38\times 10^{-3}$\\
$\rm M_{C_2}$&- &- &-\\
\hline
$\rm M_{AC}$&- &-  &0.12 \\
$\rm M_{MgSiO_3}$& $2.42\times 10^{-2}$  &- &- \\
$\rm M_{Mg_2SiO_4}$& 0.26 &$2.71\times 10^{-2}$ &-\\
$\rm M_{Fe_3O_4}$& 0.20 &-&-\\
$\rm M_{SiO_2}$& 0.82&- &-\\
$\rm M_{Al_2O_3}$&-& $1.58\times 10^{-3}$  &-\\
$\rm M_{Fe}$&0.31 &-&-  \\
\hline
$\rm M_{dust}$&1.61 &$ 2.87\times 10^{-2}$&0.12\\
\hline
\end{tabular}
\end{table*}
\begin{table*}
\caption{The total mass of dust and in each grain species after the passage of the reverse shock of increasing strength for all Pop~III SN ejecta models.}
\label{rev_core_collapse}
\begin{tabular}{@{}lcccccccc}
\hline
\multicolumn{9}{c}{Pop~III SN - Reverse Shock - Fully Mixed Ejecta Models} \\
\hline
\multicolumn{9}{c}{\bf rev1} \\
\hline
       & $\rm 13 M_{\odot}$ &  $\rm 15 M_\odot$ & $\rm 20 M_\odot$ & $\rm 25 M_\odot$  & $\rm 30 M_\odot$  &$\rm 35 M_\odot$ &$\rm 50 M_\odot$&$\rm 80 M_\odot$ \\
\hline
 $\rm M_{AC}$& $4.78\times 10^{-2}$&$9.82\times 10^{-2}$&$6.71\times 10^{-3}$&$3.59\times 10^{-2}$&0.10&-&-&-\\
 $\rm M_{MgSiO_3}$&$3.45\times 10^{-5}$&$1.19\times 10^{-5}$&-&$1.20\times 10^{-4}$&$1.58\times 10^{-3}$&-& $1.44\times 10^{-6}$&$1.68\times 10^{-4}$\\
 $\rm M_{Mg_2SiO_4}$& $7.72\times 10^{-3}$&$2.13\times 10^{-2}$&0.17& $7.45\times 10^{-2}$&$9.41\times 10^{-2}$&0.39&0.24&0.27\\
 $\rm M_{Fe_3O_4}$& $1.25\times 10^{-3}$&$3.49\times 10^{-3}$&$2.35\times 10^{-2}$&$1.21\times 10^{-2}$&$2.85\times 10^{-2}$&$5.54\times 10^{-2}$&$7.06\times 10^{-2}$& $5.77\times 10^{-2}$\\
 $\rm M_{SiO_2}$& $2.22\times 10^{-3}$&$8.59\times 10^{-3}$&$4.76\times 10^{-2}$& $6.33\times 10^{-2}$&0.15&$3.37\times 10^{-2}$&0.37&0.47\\
 $\rm M_{Al_2O_3}$&$7.29\times 10^{-5}$&$1.34\times 10^{-4}$&$6.82\times 10^{-4}$&$2.75\times 10^{-4}$&$2.89\times 10^{-4}$&$6.33\times 10^{-4}$&$1.52\times 10^{-4}$&$2.24\times 10^{-4}$\\
\hline
 $\rm M_{dust}$& $5.9\times 10^{-2}$ & $0.13$ &  $0.25$ & $0.19$ & $0.38$ &  $0.48$ &  $0.68$ & $0.80$\\
\hline
\multicolumn{9}{c}{\bf rev2} \\
\hline
       & $\rm 13 M_{\odot}$ &  $\rm 15 M_\odot$ & $\rm 20 M_\odot$ & $\rm 25 M_\odot$  & $\rm 30 M_\odot$  &$\rm 35 M_\odot$ &$\rm 50 M_\odot$&$\rm 80 M_\odot$ \\
\hline
 $\rm M_{AC}$&$2.43\times 10^{-2}$& $5.53\times 10^{-2}$&$2.83\times 10^{-3}$&$1.71\times 10^{-2}$&$5.31\times 10^{-2}$&-&-&-\\
 $\rm M_{MgSiO_3}$&$1.11\times 10^{-5}$&$3.86\times 10^{-6}$&-&$4.18\times 10^{-5}$&$5.30\times 10^{-4}$&-& $5.68\times 10^{-7}$& $5.55\times 10^{-5}$\\
 $\rm M_{Mg_2SiO_4}$&$2.48\times 10^{-3}$&$6.76\times 10^{-3}$&$7.82\times 10^{-2}$&$2.49\times 10^{-2}$&$3.01\times 10^{-2}$&0.16&$8.96\times 10^{-2}$&$8.66\times 10^{-2}$\\
 $\rm M_{Fe_3O_4}$&$5.36\times 10^{-4}$&$1.13\times 10^{-3}$&$7.07\times 10^{-3}$&$3.58\times 10^{-3}$&$8.07\times 10^{-3}$&$1.63\times 10^{-2}$&$1.98\times 10^{-2}$& $1.43\times 10^{-2}$\\
 $\rm M_{SiO_2}$&$7.72\times 10^{-4}$&$2.72\times 10^{-3}$&$1.87\times 10^{-2}$&$2.22\times 10^{-2}$&$5.61\times 10^{-2}$&$1.17\times 10^{-2}$&0.16&0.17\\
 $\rm M_{Al_2O_3}$&$3.73\times 10^{-5}$&$6.66\times 10^{-5}$&$2.30\times 10^{-4}$&$1.25\times 10^{-4}$&$1.06\times 10^{-4}$&$2.32\times 10^{-4}$&$7.13\times 10^{-5}$&$1.11\times 10^{-4}$\\
\hline
 $\rm M_{dust}$& $2.8\times 10^{-2}$ & $6.6\times 10^{-2}$ & $0.11$ & $6.8\times 10^{-2}$ & $0.15$ &$0.19$ & $0.27$ & $0.27$\\
\hline
\multicolumn{9}{c}{\bf rev3} \\
\hline
       & $\rm 13 M_{\odot}$ &  $\rm 15 M_\odot$ & $\rm 20 M_\odot$ & $\rm 25 M_\odot$  & $\rm 30 M_\odot$  &$\rm 35 M_\odot$ &$\rm 50 M_\odot$&$\rm 80 M_\odot$ \\
\hline
 $\rm M_{AC}$&$9.27\times 10^{-3}$&$2.40\times 10^{-2}$&$9.16\times 10^{-4}$&$5.80\times 10^{-3}$&$1.95\times 10^{-2}$&-&-&-\\
 $\rm M_{MgSiO_3}$&$5.11\times 10^{-6}$&$1.34\times 10^{-6}$&-&$1.28\times 10^{-5}$&$1.59\times 10^{-4}$&-&-&$1.67\times 10^{-5}$\\
 $\rm M_{Mg_2SiO_4}$&$1.17\times 10^{-3}$&$2.39\times 10^{-3}$&$2.67\times 10^{-2}$&$7.62\times 10^{-3}$&$9.55\times 10^{-3}$&$5.50\times 10^{-2}$&$2.62\times 10^{-2}$&
$2.69\times 10^{-2}$ \\
 $\rm M_{Fe_3O_4}$&$4.17\times 10^{-4}$&$7.37\times 10^{-4}$&$2.32\times 10^{-3}$&$1.81\times 10^{-3}$&$2.80\times 10^{-3}$&$5.45\times 10^{-3}$&$6.43\times 10^{-3}$& $5.91\times 10^{-3}$\\
 $\rm M_{SiO_2}$&$4.57\times 10^{-4}$&$1.11\times 10^{-3}$&$5.97\times 10^{-3}$&$7.02\times 10^{-3}$&$1.65\times 10^{-2}$&$3.87\times 10^{-3}$&$5.05\times 10^{-2}$&
$4.97\times 10^{-2}$\\
$\rm M_{Al_2O_3}$&$2.80\times 10^{-5}$&$5.11\times 10^{-5}$&$1.29\times 10^{-4}$&$8.39\times 10^{-5}$&$6.69\times 10^{-5}$&$1.52\times 10^{-4}$&$5.26\times 10^{-5}$&
$1.03\times 10^{-4}$\\
\hline
 $\rm M_{dust}$& $1.1\times 10^{-2}$ & $2.8\times 10^{-2}$ &$3.6\times 10^{-2}$ & $2.2\times 10^{-2}$ & $4.9\times 10^{-2}$& $6.4\times 10^{-2}$ & $8.3\times 10^{-2}$ & 
$8.3\times 10^{-2}$\\
\hline
\end{tabular}
\end{table*}
\begin{table*}
\caption{The total mass of dust and in each grain species after the passage of the reverse shock of increasing strength for selected unmixed Pop~III SN ejecta models.}
\label{rev_unmixed}
\begin{tabular}{@{}lccccccccc}
\hline
\multicolumn{9}{c}{Pop~III SN - Reverse Shock - Unmixed Ejecta Models} \\
\hline
\multicolumn{9}{c}{\bf rev1} \\
\hline
       & $\rm 15 M_\odot$ & $\rm 15 M_\odot$&$\rm 15 M_\odot$& $\rm 30 M_\odot$& $\rm 30 M_\odot$& $\rm 30 M_\odot$&$\rm 50 M_\odot$ &$\rm 50 M_\odot$ &$\rm 50 M_\odot$  \\
\hline
  & A & B & C & A & B &C & A & B&C \\ 
\hline
 $\rm M_{AC}$&- & $1.19\times 10^{-2}$ & $3.47\times 10^{-2}$&-&-&$3.06\times 10^{-3}$&-&-&$1.19\times 10^{-1}$\\ 
 $\rm M_{MgSiO_3}$&-&-&-&-&-&-&$2.42\times 10^{-2}$&-&-\\
 $\rm M_{Mg_2SiO_4}$&0.10 & -&-&0.21&$2.05\times 10^{-5}$&-&0.26&$1.96\times 10^{-2}$&-\\
 $\rm M_{Fe_3O_4}$& $7.45\times 10^{-2}$ & -& -&0.13&-&-&0.20&-&-\\ 
 $\rm M_{SiO_2}$&0.12 &-&-&0.56&-&-&0.82&-&-\\ 
 $\rm M_{Al_2O_3}$& $1.27\times 10^{-3}$ &-&-&-&-&-&-&$4.13\times 10^{-4}$&-\\
 $\rm M_{Fe}$&$1.20\times 10^{-2}$&-&-&-&-&-&0.31&-&-\\
\hline
 $\rm M_{dust}$&$3.16\times 10^{-1}$ &$1.19\times 10^{-2}$&$3.47\times 10^{-2}$&$8.94\times 10^{-1}$&$2.05\times 10^{-5}$&$3.06\times 10^{-3}$ &1.61&$2.0\times 10^{-2}$&$1.19\times 10^{-1}$\\
\hline
\multicolumn{9}{c}{\bf rev2} \\
\hline
  & $\rm 15 M_\odot$ & $\rm 15 M_\odot$&$\rm 15 M_\odot$& $\rm 30 M_\odot$& $\rm 30 M_\odot$& $\rm 30 M_\odot$&$\rm 50 M_\odot$ &$\rm 50 M_\odot$ &$\rm 50 M_\odot$\\
\hline
  & A & B & C & A & B &C & A & B&C \\       
\hline
 $\rm M_{AC}$&-&$1.19\times 10^{-2}$&$1.35\times 10^{-2}$&-&-&$1.11\times 10^{-3}$&-&-&$7.02\times 10^{-2}$\\
 $\rm M_{MgSiO_3}$&-&-&-&-&-&-&$2.42\times 10^{-2}$&-&-\\
 $\rm M_{Mg_2SiO_4}$&$9.53\times 10^{-2}$&-&-&$3.79\times 10^{-2}$&-&-&0.26&$6.37\times 10^{-3}$&-\\
 $\rm M_{Fe_3O_4}$&$4.69\times 10^{-2}$&-&-&$4.84\times 10^{-4}$&-&-&0.20&-&-\\
 $\rm M_{SiO_2}$&$7.22\times 10^{-2}$&-&-&0.30&-&-&0.82&-&- \\
 $\rm M_{Al_2O_3}$&$1.01\times 10^{-5}$&-&-&-&-&-&-&$4.50\times 10^{-5}$&-\\
 $\rm M_{Fe}$&$4.24\times 10^{-3}$&-&-&-&-&-&0.31&-&-\\
\hline
 $\rm M_{dust}$&$2.19\times 10^{-1}$&$1.19\times 10^{-2}$&$1.35\times 10^{-2}$&$3.37\times 10^{-1}$&-&$1.11\times 10^{-3}$&1.61 &$6.42\times 10^{-3}$&$7.02\times 10^{-2}$\\
\hline
\multicolumn{9}{c}{\bf rev3} \\
\hline
& $\rm 15 M_\odot$ & $\rm 15 M_\odot$&$\rm 15 M_\odot$& $\rm 30 M_\odot$& $\rm 30 M_\odot$& $\rm 30 M_\odot$&$\rm 50 M_\odot$ &$\rm 50 M_\odot$ &$\rm 50 M_\odot$  \\
\hline
  & A & B & C & A & B &C & A & B&C \\  
\hline
 $\rm M_{AC}$&-&$3.84\times 10^{-3}$&$2.77\times 10^{-3}$&-&-&$9.08\times 10^{-4}$&-&-&$1.61\times 10^{-2}$\\
 $\rm M_{MgSiO_3}$&-&-&-&-&-&-&$2.42\times 10^{-2}$&-&-\\
 $\rm M_{Mg_2SiO_4}$& $3.12\times 10^{-2}$ &-&-&-&-&-&0.26&$6.53\times 10^{-4}$&-\\
 $\rm M_{Fe_3O_4}$&$4.17\times 10^{-3}$ &-&-&-&-&-&0.20&-&-\\
 $\rm M_{SiO_2}$&$8.36\times 10^{-3}$ &-&-&$1.26\times 10^{-2}$&-&-&0.81&-&-\\
 $\rm M_{Al_2O_3}$&-&-&-&-&-&-&-&-&-\\
 $\rm M_{Fe}$&$1.74\times 10^{-5}$&-&-&-&-&-&0.31&-&-\\
\hline
 $\rm M_{dust}$&$4.38\times 10^{-2}$&$3.84\times 10^{-3}$&$2.77\times 10^{-3}$&$1.26\times 10^{-2}$&-&$9.08\times 10^{-4}$&1.61&$6.53\times 10^{-4}$&$1.61\times 10^{-2}$\\
\hline
\end{tabular}
\end{table*}
\begin{table*}
\caption{Properties of the faint Pop~III SN progenitors, including [Fe/H], the explosion energy [$10^{51}$~erg], the mass of the helium core, 
the ejecta velocity [$\rm km~s^{-1}$], the mass of the ejecta, the gas number density [$\rm cm^{-3}$] and 
the radius of He core, $\rm R_0 $ [cm] at $\rm t=t_{ini}$ [sec] when adiabatic expansion starts (see text); the initial masses of C, O, Mg, Si, 
Al and Fe, and the mass of molecules, CO, SiO, $\rm O_{2}$, $\rm C_{2}$ and the grains formed in the expanding 
ejecta [$\rm M_{\odot}$].}
\label{tabappIII:faint}
\begin{tabular}{@{}lcccc}
\hline
\multicolumn{5}{c}{Pop~III Faint SN - Fully Mixed Ejecta Models} \\
\hline
     & HE1327-2326 &  HE0107-5240 & HE0557-4840 & SMSSJ031300 \\
\hline
 [Fe/H] &  $-5.76$   & $-5.54$   &  $-4.81$   & $-7.1$ \\    
\hline
     & $\rm 30 M_{\odot}$ &  $\rm 35 M_\odot$ & $\rm 20 M_\odot$ & $\rm 80 M_\odot$\\  
\hline
 $\rm E_{exp} $ & $1.6$ & $1.3$ & $1.0$ & $5.2$ \\
 $\rm M_{\rm He_{core}}$& $7.22$ &$11.13$ &$5.18$&$31.94$\\
 $\rm \rm{v_{eje}}$&$3273$&$2833$&$3301$&$3943$\\
 $\rm M_{eje}$&$25$&$27$&$15$&$56$ \\
 $\rm M_{cut}$&$5$& $8$ & $5$& $24$\\
\hline
 $\rm R_0$&$4.68\times 10^{14}$&$1.98\times 10^{14}$&$1.47\times 10^{14}$& $4.78\times 10^{14}$\\
 $\rm t_{ini}$&$3.96\times 10^{6}$&$3.11\times 10^{6}$&$1.58\times 10^{6}$& $4.14\times 10^{6}$\\
 $\rm n_0$& $3.27\times 10^{11}$&$1.36\times 10^{12}$&$2.46\times 10^{10}$&$5.17\times 10^{11}$\\
\hline
 $\rm M(^{56}Ni)$&$3.68\times 10^{-7}$&$9.12\times 10^{-7}$&$9.19\times 10^{-7}$&$1.43\times 10^{-7}$\\
 $\rm M_{C}$ & $0.51$& $0.297$&$2.76\times 10^{-3}$&$0.887$\\
 $\rm M_{O}$ & $1.19$& $0.195$& $6.37\times 10^{-4}$&$1.973$ \\
 $\rm M_{Mg}$ & $9.74\times 10^{-5}$& $3.02\times 10^{-5}$&$3.37\times 10^{-5}$&$3.26\times 10^{-3}$ \\
 $\rm M_{Si}$ & $2.81\times 10^{-5}$& $2.07\times 10^{-5}$&$3.78\times 10^{-5}$&$5.19\times 10^{-6}$ \\
 $\rm M_{Ca}$ &$ 1.50\times 10^{-6}$& $1.50\times 10^{-6}$&$2.21\times 10^{-6}$&$3.82\times 10^{-7}$ \\
 $\rm M_{Fe}$ & $1.58\times 10^{-5}$& $2.24\times 10^{-6}$&$2.82\times 10^{-5}$&$4.7\times 10^{-6}$\\
\hline
 $\rm M_{CO}$ &$0.74$&$0.342$&$1.12\times 10^{-4}$ & $1.44$ \\ 
 $\rm M_{SiO}$&$4.0\times 10^{-5}$&$2.7\times 10^{-5}$&$1.8\times 10^{-7}$&$3.5\times 10^{-6}$\\
 $\rm M_{O_2}$&$1.2\times 10^{-5}$&-&-& $2.4\times 10^{-5}$\\
 $\rm M_{C_2}$&-&$2.6\times 10^{-2}$&-&-\\
\hline
 $\rm M_{AC}$& $0.196$&$0.124$ & $2.71\times 10^{-3}$ &$0.269$ \\
\hline
\end{tabular}
\end{table*}
\begin{table*}
\caption{Properties of the faint Pop~III SN unmixed models, including the explosion energy [$10^{51}$~erg], the mass of the ejecta, 
the initial masses of C, O, Mg, Si, Al and Fe, and the mass of CO and the grains formed in the expanding ejecta [$\rm M_{\odot}$]. 
Each model name identifies the progenitor mass.}
\label{tabappIII:faint_unmixed}
\begin{tabular}{@{}l|cc|cc|cc|cc}
\hline
\multicolumn{7}{c}{Pop~III Faint SN - Unmixed Ejecta Models} \\
\hline
     & HE1327-2326 & HE1327-2326 &  HE0107-5240 & HE0107-5240 & SMSSJ031300& SMSSJ031300 \\
\hline
  &$\rm 30 M_{\odot}$&$\rm 30 M_{\odot}$& $\rm 35 M_{\odot}$&$\rm 35 M_\odot$&$\rm 80 M_{\odot}$&$\rm 80 M_\odot$\\  
\hline
  & A & B & A & B  & A & B \\ 
\hline
 $\rm E_{exp} $ & $1.6$ & $1.6$ & $1.3$&$1.3$ & $5.2$ & $5.2$ \\
 $\rm M_{eje} $ & $25$ & $25$ & $27$& $27$&$56$& $56$ \\
 $\rm v_{eje}$&$3273$&$3273$& $2833$& $2833$&$3943$& $3943$\\
 $\rm M_{cut}$&$5$& $5$ & $8$& $8$ & $24$& $24$\\
\hline
 $\rm R_0$&$2.18\times 10^{14}$&$4.73\times 10^{14}$&$1.45\times 10^{14}$&$2.21\times 10^{14}$& $2.29\times 10^{14}$ &$4.57\times 10^{14}$\\          
 $\rm t_{ini}$&$1.90\times 10^{6}$&$2.45\times 10^{6}$&$2.66\times 10^{6}$&$1.84\times 10^{6}$& $2.24\times 10^{6}$& $4.14\times 10^{6}$\\        
 $\rm n_0$&$1.76\times 10^{13}$&$2.4\times 10^{9}$&$1.98\times 10^{12}$&$2.96\times 10^{9}$&$2.33\times 10^{13}$&$3.31\times 10^{11}$\\        
\hline
 $\rm M(^{56}Ni)$&$3.68\times 10^{-7}$&-&$9.12\times 10^{-7}$&-&$1.43\times 10^{-7}$&-\\
 $\rm M_{C}$&$0.508$ &$6.35\times 10^{-3}$& $0.296$&$7.61\times 10^{-4}$&$0.547$&$0.339$\\ 
 $\rm M_{O}$&$1.162$ &$2.60\times 10^{-2}$& $0.192$&$3.77\times 10^{-3}$&$1.948$&$0.024$ \\
 $\rm M_{Mg}$&$9.74\times 10^{-5}$& - &$3.00\times 10^{-5}$&-&$3.26\times 10^{-3}$&-\\    
 $\rm M_{Si}$&$2.81\times 10^{-5}$& - &$2.06\times 10^{-5}$&-&$5.19\times 10^{-6}$&-\\
 $\rm M_{Ca}$&$1.50\times 10^{-6}$& - &$1.50\times 10^{-6}$&-&$3.82\times 10^{-7}$&-\\ 
 $\rm M_{Fe}$&$1.58\times 10^{-5}$& - &$2.23\times 10^{-5}$&-&$4.7\times 10^{-6}$&-\\
\hline
 $\rm M_{CO}$& $1.18$ &$3.80\times 10^{-4}$ &$0.33$ &$3.21\times 10^{-5}$&$1.276$& $0.043$\\
 $\rm M_{SiO}$&-&-&- &-&-&-\\
 $\rm M_{O_2}$&$0.11$&-&- &-& $0.513$&-\\
 $\rm M_{C_2}$&-&-&$9.05\times 10^{-2}$&-&-&$0.21$ \\
\hline
 $\rm M_{AC}$ & - & $6.18\times 10^{-3}$ & $6.21\times 10^{-2}$ &$7.47\times 10^{-4}$ &- &$0.112$\\
\hline
\end{tabular}
\end{table*}
\begin{table*}
\caption{The total mass of dust after the passage of the reverse shock of increasing strenght for all faint Pop~III SN ejecta models.}
\label{rev_faint}
\begin{tabular}{@{}lcccc}
\hline
\multicolumn{5}{c}{Pop~III Faint SN - Reverse-Shock - Fully Mixed Ejecta Models} \\
\hline
     & HE1327-2326 &  HE0107-5240 & HE0557-4840 & SMSSJ031300 \\
\hline
     & $\rm 30 M_{\odot}$ &  $\rm 35 M_\odot$ & $\rm 20 M_\odot$ & $\rm 80 M_\odot$\\  
\hline
\multicolumn{5}{c}{\bf rev1} \\
\hline
 $\rm M_{AC}$& $0.11$&0.10& $8.3\times 10^{-4}$&0.15\\
\hline
\multicolumn{5}{c}{\bf rev2} \\
\hline
 $\rm M_{AC}$&$5.7\times 10^{-2}$&$6.2\times 10^{-2}$&$3.0\times 10^{-4}$&$7.8\times 10^{-2}$\\
\hline
\multicolumn{5}{c}{\bf rev3} \\
\hline
 $\rm M_{AC}$&$1.99\times 10^{-2}$&$2.8\times 10^{-2}$& $9.0\times 10^{-5}$&$2.8\times 10^{-2}$\\
\hline
\end{tabular}
\end{table*}
\begin{table*}
\caption{The total mass of dust after the passage of the reverse shock of increasing strenght for selected unmixed faint Pop~III SN ejecta models.}
\label{rev_faint_unmixed}
\begin{tabular}{@{}l|cc|cc|cc|cc}
\hline
\multicolumn{7}{c}{Pop~III Faint SN - Reverse-Shock -  Unmixed Ejecta Models} \\
\hline
     & HE1327-2326 & HE1327-2326 &  HE0107-5240 & HE0107-5240 & SMSSJ031300& SMSSJ031300 \\
\hline
  &$\rm 30 M_{\odot}$&$\rm 30 M_{\odot}$& $\rm 35 M_{\odot}$&$\rm 35 M_\odot$&$\rm 80 M_{\odot}$&$\rm 80 M_\odot$\\  
\hline
  & A & B & A & B  & A & B \\ 
\hline
\multicolumn{7}{c}{\bf rev1} \\
\hline
 $\rm M_{AC}$& -&  $7.54\times 10^{-4}$& $6.20\times 10^{-2}$&$5.88\times 10^{-5}$&-&$7.19\times 10^{-2}$\\                   
\hline
\multicolumn{7}{c}{\bf rev2} \\
\hline
 $\rm M_{AC}$& -& $1.54\times 10^{-4}$&$6.20\times 10^{-2}$ & $1.54\times 10^{-5}$  &- &$1.39\times 10^{-2}$\\
\hline
\multicolumn{7}{c}{\bf rev3} \\
\hline
 $\rm M_{AC}$&-& $1.02\times 10^{-4}$ & $6.20\times 10^{-2}$&-&-&$8.65\times 10^{-3}$\\
\hline
\end{tabular}
\end{table*}
\label{lastpage}
\end{document}